\documentclass[twocolumn,colorlinks=true,linkcolor=blue,urlcolor=blue,citecolor=magenta]{svjour3}

\usepackage[T1]{fontenc}
\usepackage[utf8]{inputenc}

\usepackage{amsmath} 
\usepackage{booktabs}
\usepackage{pifont} 
\usepackage{hyperref}
\usepackage{fancyvrb}
\usepackage{graphicx}
\usepackage{xcolor} 

\DeclareMathOperator*{\argmax}{arg\,max}

\renewcommand{\vec}[1]{\mathbf{#1}}

\newcommand{\LLM}{\textrm{LLM}}
\newcommand{\cmark}{\ding{51}}

\newcommand{\hi}[1]{\vspace{.25em}\noindent {\textbf{#1}}}

\begin{document}

\title{A Survey of LLM Inference Systems}

\author{James Pan \and Guoliang Li}

\institute{
James Pan \at
Department of Computer Science and Technology \\
Tsinghua University, Beijing, China \\
\email{jpan@tsinghua.edu.cn}
\and
Guoliang Li \at
Department of Computer Science and Technology \\
Tsinghua University, Beijing, China \\
\email{liguoliang@tsinghua.edu.cn}
}

\date{}

\maketitle

\setcounter{tocdepth}{3}

\begin{abstract}
The past few years has witnessed specialized large language model (LLM) inference systems, such as vLLM, SGLang, Mooncake, and DeepFlow, alongside rapid LLM adoption via services like ChatGPT. Driving these system design efforts is the unique autoregressive nature of LLM request processing, motivating new techniques for achieving high performance while preserving high inference quality over high-volume and high-velocity workloads. While many of these techniques are discussed across the literature, they have not been analyzed under the framework of a complete inference system, nor have the systems themselves been analyzed and compared.

In this survey, we review these techniques, starting from operators and algorithms for request processing, then moving on to techniques for model optimization and execution, including kernel design, batching, and scheduling, before ending with techniques for memory management, including paged memory, eviction and offloading techniques, quantization, and cache persistence. Through these discussions, we show that these techniques fundamentally rely on load prediction, adaptive mechanisms, and cost reduction in order to overcome the challenges introduced by autoregressive generation and achieve the goals of the system. We then discuss how these techniques can be combined to form single-replica and multi-replica inference systems, including disaggregated inference systems that offer more control over resource allocation and serverless systems that can be deployed over shared hardware infrastructure. We end with a discussion of remaining challenges.
\end{abstract}

\section{Introduction}

Since the shift from recurrent neural networks to transformers for sequence generation \cite{vaswani2017attention}, large language models (LLMs) have now reached a level of quality that allow them to be used for a variety of tasks, including interactive generic question-answering \cite{beurerkellner2023lmql}, document summarization and classification \cite{sun2025hygen}, language translation \cite{freitag2017beam}, code generation \cite{chen2021evaluating}, data wrangling \cite{narayan2022can}, and unstructured data analysis \cite{wang2025aop}. This breakthrough has led to an exponential rise in LLM adoption across industry and ordinary consumers alike via services like ChatGPT, Gemini, Claude, Grok, Kimi, and DeepSeek, putting pressure on designing systems that can support high-performance model serving.

To meet this demand, specialized LLM inference systems have been developed to manage all aspects of model execution, including the basic LLM inference workflow in addition to system-level aspects such as load balancing, job batching, job scheduling, and memory management, following in the footsteps of other high-volume and high-velocity data processing systems. But the unique autoregressive nature of transformer-based LLM inference means that new techniques have had to be developed for each of these aspects.

\begin{figure}[!t]
\centering
\includegraphics[width=0.48\textwidth]{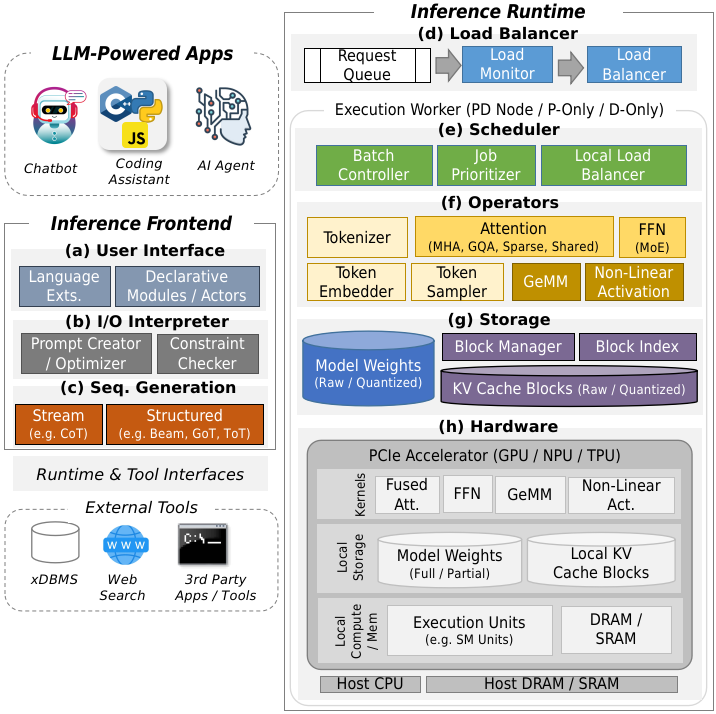}
\caption{The LLM inference stack.}
\label{fig:system-overview}
\end{figure}

Traditional data processing systems process a request by executing a sequence of operators once over the input, but LLMs need to do this a number of times that is proportional to the length of the request output. All requests take the form of a text string, and the output length depends non-deterministically on the textual contents of the string. Since the string contents can be anything a user desires, there is no such thing as typical output length. The request processing cost, especially the memory cost, therefore is unknown and can vary across an extremely wide range, even between requests with similar contents\footnote{\textit{E.g.} chain-of-thought prompting.}.

The fundamentally non-deterministic nature of the outputs leads to three key challenges for LLM inference systems. \textit{(1}) Despite recent breakthroughs, the quality of the output, in terms of how well it satisfies a task expressed by a request, still cannot be guaranteed because the output is generated by non-deterministic sampling instead of being constructed analytically from data. \textit{(2)} The indetermined number of execution rounds means that the final memory usage for processing a request is unknown, making it challenging to allocate memory for systems that handle multiple requests at a time. \textit{(3)} Likewise, since the time it takes to process a request is also indetermined, designing batching and scheduling techniques that can avoid issues like stragglers and head-of-line blocking is also challenging.

To address these challenges, LLM inference systems employ a number of techniques, spanning the user-facing frontend as well as the execution runtime, as shown in Figure~\ref{fig:system-overview}. To increase LLM quality, inference systems support a range of sequence generation techniques, such as beam search, tree-of-thoughts, graph-of-thoughts, and self-consistency, that increase the likelihood of producing high-quality output (Figure~\ref{fig:system-overview}(c)), in addition to a number of prompting techniques. The variety of techniques leads to frontend designs that aim to simplify user interactions (Figure~\ref{fig:system-overview}(a)), with features that include automatic prompt optimization and constrained output generation (Figure~\ref{fig:system-overview}(b)) aimed at reducing the burden of crafting prompts and coordinating complex workflows. To adapt to dynamic memory needs, inference systems rely on page-based block memory allocation combined with cache persistence and quantization techniques that serve to reduce the overall memory usage (Figure~\ref{fig:system-overview}(g)). To adapt to dynamic request lifecycles, inference systems use dynamic job scheduling, dynamic batching, and flexible load balancing (Figure~\ref{fig:system-overview}(d, e)) based on load prediction mechanisms, along with specialized operators and kernel implementations to reduce the overall inference cost (Figure~\ref{fig:system-overview}(f, h)).

In this survey, we discuss these techniques within the framework of a comprehensive inference system. In Section~\ref{sec:processing}, we discuss operators and sequence generation algorithms that are fundamental to performing high-quality LLM inference. In Section~\ref{sec:execution}, we discuss techniques for batching and scheduling, in addition to kernel design that seeks to develop efficient operator implementations for exploiting specialized hardware. In Section~\ref{sec:memory}, we discuss techniques for memory management, including page-based memory, eviction and offloading techniques for supporting request preemptions and long contexts, quantization, and cache persistence techniques, including techniques for cache reconstruction. Following these discussions, in Section~\ref{sec:systems} we discuss how these techniques are combined to form the current landscape of LLM inference systems, including single-replica systems aimed at environments that host a single copy of the LLM as well as multi-replica systems aimed at processing requests over multiple LLM copies. These systems in particuar allow for disaggregated architectures that allow for more control over how hardware resources are allocated.

\hi{Related Surveys.}
While many of the presented techniques have been discussed in prior surveys, in this survey we contextualize them under a comprehensive inference framework as well as elaborate on how they can be combined into dedicated LLM inference systems. In \cite{khoshnoodi2024comprehensive,li2024llm,xu2025resource,zhou2024survey}, techniques including sparse attention, MoE, decoding strategies, KV cache management, and quantization are discussed in isolation, outside of a system framework.  In \cite{li2025survey}, many of these techniques are discussed from the perspective of KV cache management. Other surveys, such as \cite{chavan2024faster}, focus on model architecture, including techniques for model pruning and knowledge distillation as well as quantization, or focus on techniques for improving inference quality \cite{kim2024trustworthy}.

\section{Request Processing}
\label{sec:processing}

A transformer-based language model operates over a discrete finite set of tokens, producing one output token in response to a sequence of input tokens, $x_1\dots x_p$, all being members of this set. This is done by mapping the token set onto a high-dimensional vector space, allowing simple linear transformations to be applied to the token embeddings to produce a \textit{contextualized} embedding for $x_p$ from which the output token is ultimately selected.

All LLM inference systems that will be presented in Section~\ref{sec:systems} use transformer-based models that follow the same basic inference workflow, principally relying on attention, feed-forward, and token sampling operators over the embeddings to produce output tokens (Section~\ref{sec:workflow}). Nevertheless, the high fundamental costs of attention and feed-forward operators, along with the need to produce tokens that lead to semantically coherent and otherwise acceptable text\footnote{From a user perspective, the input and output is text, not tokens. Behind the scenes, user text is deterministically converted into tokens by an LLM application, \textit{e.g.} a chatbot, before being fed into the LLM. Output tokens are likewise converted back into text before being presented to the user. Tools like Tiktokenizer (\url{https://github.com/dqbd/tiktokenizer}) can be used to view the tokens corresponding to a text.}, has inspired much effort on operator design (Section~\ref{sec:operators}). Moreover, due to the inherent limitation that tokens can only be generated one at a time, generating long textual sequences involves multiple rounds of model execution, and as each output token can be highly sensitive to the given input for each round, these same concerns have led additionally to various techniques for sequence generation (Section~\ref{sec:generation}). Faced with this diverse landscape of request processing techniques, system designers must carefully balance the various advantages and disadvantages of these techniques with respect to how the rest of the system is designed, and the downstream consequences are discussed in Section~\ref{sec:processing-discussion}.

\subsection{Inference Workflow}
\label{sec:workflow}

A request consists of an initial input, $x_1\dots x_p$, called the prompt or prefix. The response is a completed sequence, $x_1\dots x_p\dots x_n$, based on the prompt. Computing token $x_{i+1}$, for $i\geq p$, requires one execution of the model over all the previous tokens, \textit{i.e.} $x_{i+1} = \LLM(x_1\dots x_i)$, and so the output sequence is formed one token at a time by feeding back previous tokens, a process called autoregressive generation.

\begin{figure}[!t]
\centering
\includegraphics[width=.48\textwidth]{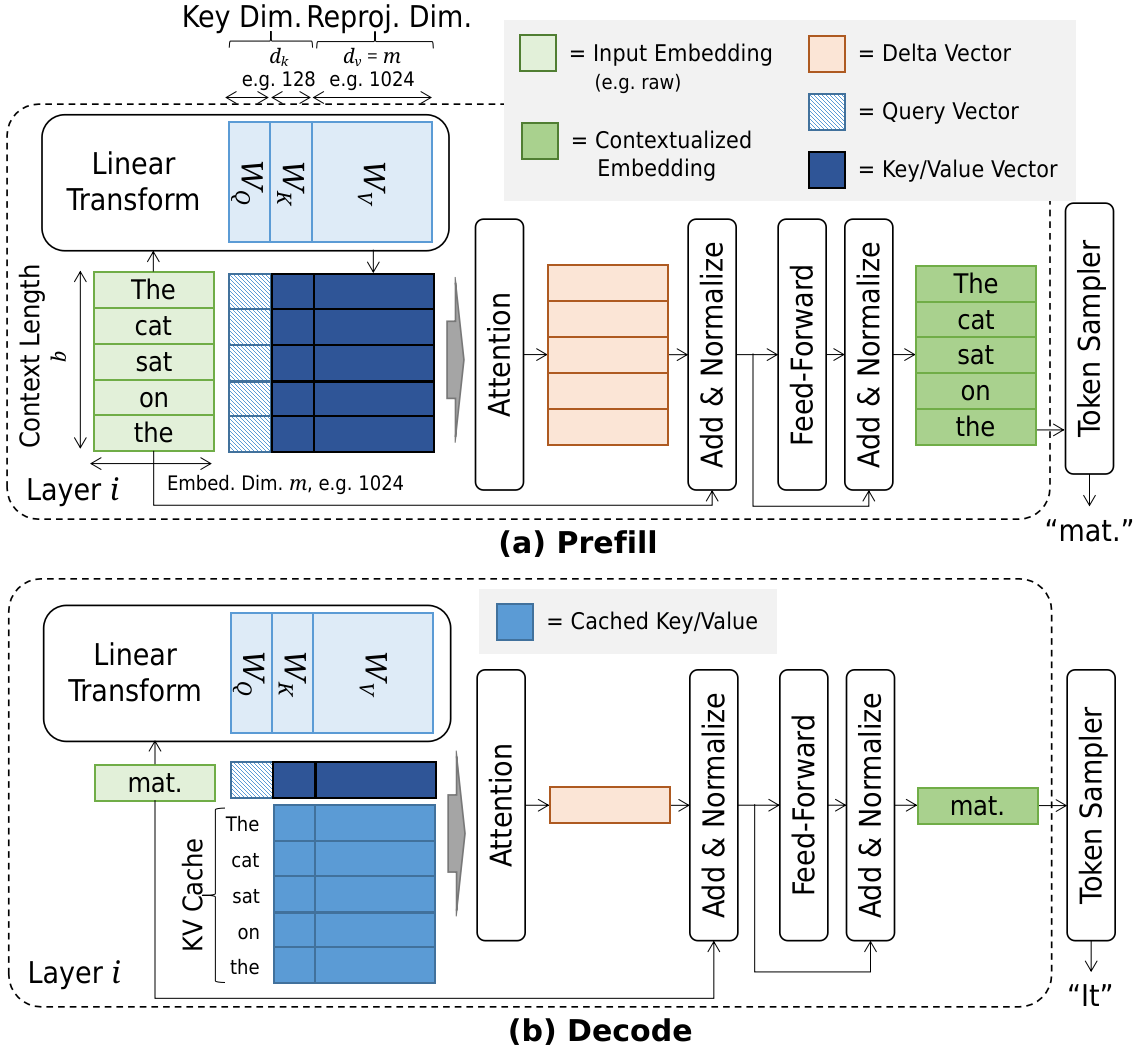}
\caption{Prefill (\textbf{a}) and decode (\textbf{b}) in one transformer layer.}
\label{fig:workflow}
\vspace{-1em}
\end{figure}

To produce one output token, each input token is mapped onto a high-dimensional embedding via an embedding model\footnote{Including a positional embedding sub-module.}. The embeddings then flow through a series of transformer layers, each functionally identical but parameterized differently, to become contextualized. The contextualized embedding for the last input token of the input sequence is fed into the token sampler to produce the output token. Figure~\ref{fig:workflow} introduces the workflow.

Inside each transformer layer is an attention operator coupled with a feed-forward network (FFN), with a normalization operator in between \cite{vaswani2017attention}. The attention operator and the FFN work mainly by multiplying embeddings by various weight matrices, the values of which are determined through training. The fundamental compute and storage (memory) costs of these operators therefore depends on the size of these matrices. For the attention operator specifically, applying the operator to the $i$th embedding requires linear projections (\textit{i.e.} the \textit{key} and \textit{value} vectors) of the $j$th embeddings, $j<i$. Since sequence generation is autoregressive, the number of these projections (\textit{i.e.} the size of the \textit{KV cache}) grows during the lifetime of the request. Hence, the total compute and memory costs for processing a request depends on the length of its output.

\hi{Performance and Accuracy.}
High cost manifests as high end-to-end latency, encompassing the time between when the request is submitted and when the full output sequence is generated. But because of the autoregressive nature of LLMs, the output can be streamed to the user as the tokens are generated. Consequently, for interactive applications like question-answering, long latency may not necessarily lead to a poor user experience as long as the time between tokens (TBT) and time to first token (TTFT) are short. Other applications, such as document summarization, may be more sensitive to token or request throughput. For requests that are sensitive to both latency and throughput, goodput may be a more useful performance measure, defined as the number requests completed within a certain latency target (\textit{i.e.} service-level objective, or SLO), per second \cite{wang2024revisiting}.

Users also demand outputs that are high quality. But since a user prompt can express a huge variety of tasks, ranging from knowledge retrieval to semantic operations \cite{wang2025aop}, the quality of the output depends highly on the task, precluding a universal quality indicator. For measuring the general quality of an LLM, perplexity is commonly used \cite{chen1998evaluation}. Otherwise, metrics such as BLEU \cite{papineni2002bleu} and ROUGE \cite{lin2004rouge} have been used for specific tasks, such as machine translation and document summarization, respectively.

\subsection{Operators}
\label{sec:operators}

\begin{figure}[!t]
\centering
\includegraphics[width=0.45\textwidth]{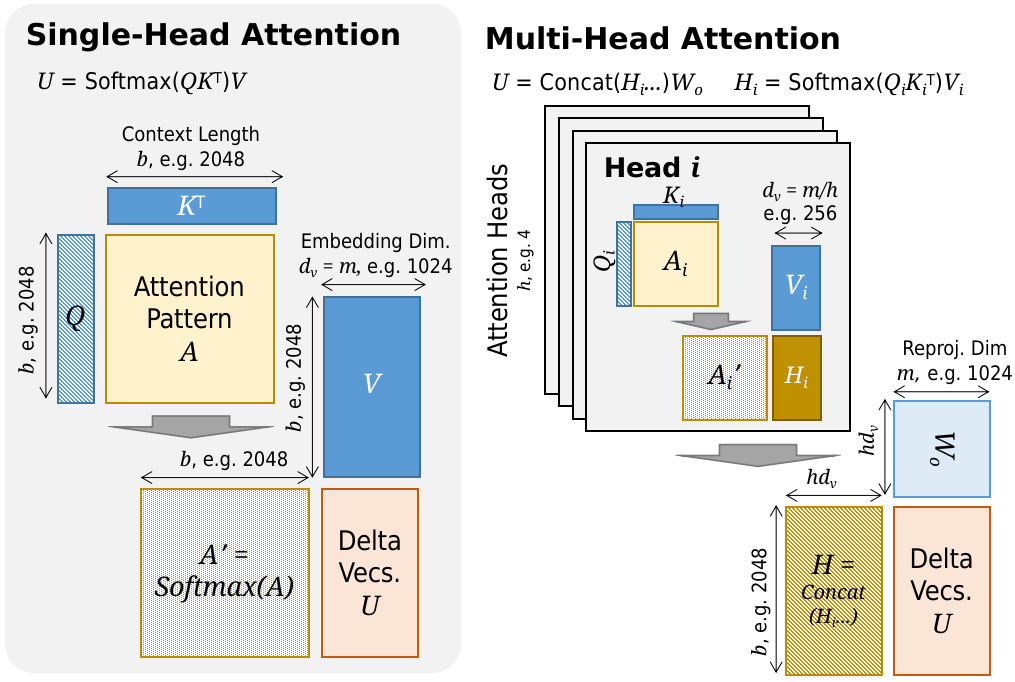}
\caption{Single-headed and multi-headed attention. In MHA, the low-dimensional delta vectors produced by each attention head are concatenated and multiplied by a reprojection matrix.}
\label{fig:attention}
\vspace{-1em}
\end{figure}

The desire for high performance and high accuracy inference inspires new designs for the attention, FFN, and token sampling operators, beyond those in \cite{vaswani2017attention}. For the attention operator, \cite{vaswani2017attention} introduces a multi-headed attention (MHA) mechanism to improve inference quality. Grouped query attention (GQA) \cite{ainslie2023gqa} and multi-query attention (MQA) \cite{shazeer2019fast} take this idea further, lowering the fundamental compute and memory costs by reducing the size and number of transformation matrices. Likewise, many works focus on sparse attention, \textit{e.g.} \cite{wang2024flash}, where the attention calculations are only partially applied, reducing costs still further. There are also techniques for shared attention, \textit{e.g.} \cite{ye2024chunkattention}, where certain attention outputs are shared across multiple requests, thus reducing the memory burden.
For the FFN, mixture-of-experts (MoE) techniques work by splitting the FFN into a number of small independent networks \cite{liu2025survey}, thus reducing the compute burden.
For token sampling, stochastic sampling techniques, such as top-$k$ sampling and nucleus sampling \cite{holtzman2020curious}, increase inference quality by exploring beyond narrow regions of the token space, while speculative decoding techniques increase token throughput by taking advantage of small draft models, followed by parallel verification \cite{yan2025decoding}.

\subsubsection{Attention}

Given the embedding $\vec{x}_i$ for token $x_i$, the attention operator produces a delta vector $\vec{u}_i$ such that the contextualized embedding vector, $\vec{x}_i\gets\vec{x}_i+\vec{u}_i$, better captures the semantic meaning of the token\footnote{To give a simplified example, the word \textit{plant} can be either a noun or a verb. Each meaning occupies a distinct coordinate in the embedding space, but token embedding $\vec{x}_i$ can only point to one of the coordinates at a time. The delta vector $\vec{u}_i$, produced by the attention operator, is used to update $\vec{x}_i$ so that it points to the intended meaning. This vector is calculated by examining the other tokens surrounding $x_i$, in other words the context.}. Vector $\vec{u}_i$ is calculated using
\begin{equation}
\label{eq:decode-attention}  \vec{u}_i=\textrm{softmax}\left(\vec{q}_i^\top K^\top\right)V,
\end{equation}
where $\vec{q}_i^\top =\vec{x}_i^\top W_q$. Vector $\vec{q}_i$ is called the query of token $x_i$, and $\vec{q}_i^\top K^\top$ is called the attention pattern.

The $K$ and $V$ matrices are formed from tokens preceding $x_i$, \textit{i.e.} the tokens $x_1\dots x_{i-1}$. Specifically, the $j$th row or \textit{key} of $K$, or $\vec{k}_j$, and the $j$th row or \textit{value} of $V$, or $\vec{v}_j$, correspond to linear projections\footnote{The weight matrices $W_q$, $W_k$, and $W_v$ are learned during training and are the same for every request.} of the $j$th embedding vector, in other words $\vec{k}_j^\top = \vec{x}_j^\top W_k$ and $\vec{v}_j^\top = \vec{x}_j^\top W_v$, for all $j$ between $0$ and $i$. As a result, the $K$ and $V$ matrices grow by one row for every execution of the attention operator in order to incorporate the projections of the $i$th token. Since the $K$ and $V$ matrices are used across the life of a request, they are persisted in memory and are commonly referred to as the \textit{KV cache}.

The attention operator can also produce delta vectors for a whole sequence of tokens in a single invocation by vertically concatenating the embeddings into an embeddings matrix and replacing softmax with row-wise softmax. The key and value vectors for each of the embeddings are calculated all at once using the embeddings matrix before calculating the attention pattern for the entire matrix. This capability leads to two phases of request processing, the prefill phase which applies this technique to the prompt tokens  (Figure~\ref{fig:workflow}(a)), and the decode phase which applies Equation~\ref{eq:decode-attention} to the output tokens as they are generated (Figure~\ref{fig:workflow}(b)). Moreover, as long as the correct key and value vectors are used, the embeddings matrix can even contain embeddings from multiple independent requests, leading to batched processing (Section~\ref{sec:batching}). But as the KV cache grows, the operator becomes increasingly costly, leading to attention variants that aim to reduce the cost.

\hi{Prefill and Batching.}
For the prefill phase, all the prompt token embeddings are vertically concatenated to from a single input matrix, $X$, that is then multipled by each of the weight matrices to obtain $Q=XW_q$, $K=XW_k$, and $V=XW_v$. The update vectors are then obtained by
\begin{equation}
\label{eq:prefill-attention}  U=\textrm{softmax}\left(QK^\top\right)V,
\end{equation}
where softmax is applied row-wise\footnote{In Equations \ref{eq:decode-attention} and \ref{eq:prefill-attention}, the input to softmax sometimes appears as scaled by $\sqrt{d}$, or the square root of the projection dimension (\textit{i.e.} widths of $K$ and $Q$). The softmax itself can be replaced by other activation functions, such as sigmoid \cite{ramapuram2025}.}. The contextualized embedding vectors are obtained by $X\gets X+U$.

On data-parallel hardware, \textit{e.g.} GPUs, calculating $X\gets X+U$ by applying Equation \ref{eq:prefill-attention} once is more efficient compared to calculating it via multiple applications of Equation \ref{eq:decode-attention}, even though the fundamental costs are the same. Aside from prefill, this property also motivates request batching (Section~\ref{sec:batching}), effectively allowing multiple decodes to be performed for multiple requests in a single invocation (Figure~\ref{fig:batched-transformer}). This property is also what makes speculative decoding practical.

\begin{figure}[!t]
\centering
\includegraphics[width=0.48\textwidth]{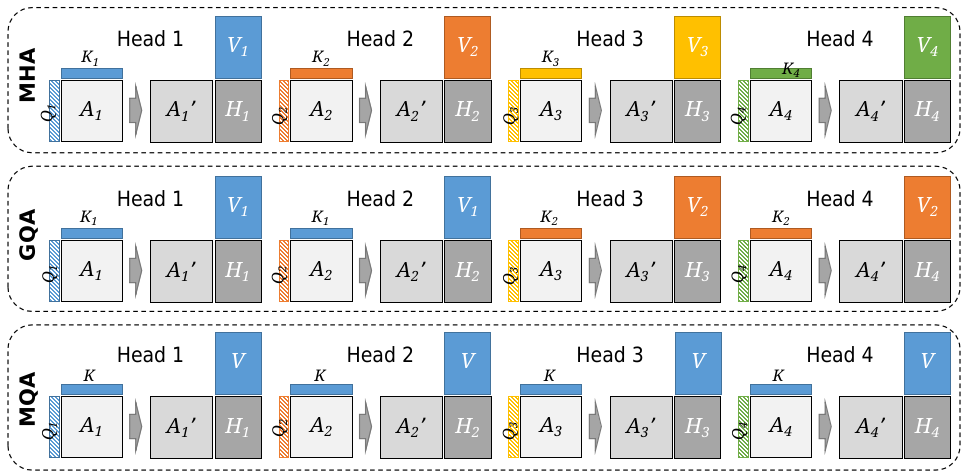}
\caption{Query grouping across 4 heads. Standard MHA uses 4 KV projections, one per head, while GQA shares a small number (in this case, 2) of KV projections across heads and MQA uses 1 projection for all the heads.}
\label{fig:attention-groups}
\vspace{-1em}
\end{figure}

\hi{Attention Variants.}
For requests with long prompts\footnote{Note that some applications, such as RALM chatbots \cite{asai2023}, may insert a large number of hidden tokens into the input that are not given by nor visible to the user, making the prompt longer than it appears.\label{ft:long-contexts}}, or for long-running requests that require a large number of decode rounds, the number of rows in $K$ and $V$ can ultimately amount to tens or even hundreds of thousands of rows, increasing both memory and compute costs. \textit{(1) Multi-headed attention} (MHA) distributes the costs by parallelizing Equations \ref{eq:decode-attention} and \ref{eq:prefill-attention} through vertically splitting $W_q$, $W_k$, and $W_v$, calculating partial delta vectors based on these mini-matrices, and then aggregating the results, while also improving inference quality \cite{vaswani2017attention}. \textit{(2) Attention grouping}, which includes grouped query attention (GQA) \cite{ainslie2023gqa} and multi-query attention (MQA) \cite{shazeer2019fast}, directly reduces the costs by dropping a number of these mini-matrices, see Figure~\ref{fig:attention-groups} and Table~\ref{tab:attention}. \textit{(3) Sparse attention} reduces the compute cost by calculating the attention pattern only for certain keys (Figure~\ref{fig:attention-masks}). Techniques for discovering these significant keys include static or query-dependent masks and filters \cite{ainslie2020etc,beltagy2020longformer,dong2024flex,jiang2024minference,lee2024infinigen,ribar2024sparq,wang2024flash,xiao2024infllm,xiao2024duoattention} and $k$-nearest neighbor search indexes \cite{chen2024magic,liu2024retrieval,zhang2024pqcache}. \textit{(4) Shared attention} reduces memory and compute costs by sharing attention patterns across multiple inputs, which could belong to a single request (\textit{e.g.} beam search) or across multiple requests (\textit{e.g.} system prompt) \cite{juravsky2024hydragen,yan2021fastseq,ye2024chunkattention,zheng2024sglang,zhu2024relay}. This technique reduces compute cost by calculating the attention pattern of the shared tokens only once while also reducing memory cost by sharing memory entries across the multiple inputs.

\begin{table}[t]
\caption{Characteristics of main attention variants.}
\label{tab:attention}
\centering
\begin{tabular}{rlccc}
\toprule
Type
  & Attention Variant
  & Lat.
  & Mem.
  & Acc. \\
\midrule
Optimization
  & Shared KVs
  & $\downarrow$
  & $\downarrow$
  & \\
  & Sparse Attention
  & $\downarrow$
  & $\downarrow$
  & $\downarrow$ \\
Structural
  & MHA \cite{vaswani2017attention}
  &
  &
  & $\uparrow$ \\
  & GQA \cite{ainslie2023gqa}
  & $\downarrow$
  & $\downarrow$
  & \\
  & MQA \cite{shazeer2019fast}
  & $\downarrow\downarrow$
  & $\downarrow\downarrow$
  & $\downarrow$ \\
\bottomrule
\end{tabular}
\end{table}

\subsubsection{Feed-Forward Network}

Given embedding $\vec{x}_i$, the feed-forward network calculates $\vec{x}_i\gets f_2(g(f_1(\vec{x}_i)))$, where $f_1(\vec{x}_i)=\vec{x}_i^\top W_1+\vec{b}_1$, $f_2(\vec{h})=\vec{h}^\top W_2+\vec{b}_2$, and $g(\vec{h})$ is a non-linear activation function, \textit{e.g.} $g(\vec{h})=\textrm{max}(0,\vec{h})$, that is applied element-wise. Substituting yields\footnote{Like the attention parameters, $W_1$, $W_2$, $\vec{b}_1$, and $\vec{b}_2$ are learned during training and are the same for every request.}
\begin{equation}
  \textrm{FFN}(\vec{x}_i) = \textrm{max}(0, \vec{x}_i^\top W_1 + \vec{b}_1)^\top W_2 + \vec{b}_2.
\end{equation}
The compute and memory costs are proportional to the sizes of the parameters. But unlike attention, where the costs vary depending on the size of the KV cache, the FFN cost is the same for all inputs.

\hi{Mixture-of-Experts.}
In an MoE system \cite{shazeer2017}, the FFN is replaced with a set of smaller networks, each called an expert. A gate determines which expert is used to process a given input. As each expert is much smaller than the original FFN, the compute cost is substantially reduced. Much of the work on MoE systems focuses on architecture of the experts, gate design, dynamic loading of experts, and hardware optimizations for expert computation. In distributed settings, expert placement and load balancing also need to be considered. We refer interested readers to \cite{liu2025survey}.

\subsubsection{Token Sampling}

Given contextualized embedding $\vec{x}_i$, the token sampler calculates $\vec{p}=\textrm{softmax}(\vec{x}_i^\top W_b)$ to yield the probability masses, or logits, across the token range\footnote{Like other parameters, $W_b$ is learned during training.}. The length of $\vec{p}$ equals the number of possible tokens, and a mapping between indices of $\vec{p}$ and the token range is used to decide the token based on a selected index.

Greedily selecting the maximum logit index at each decoding round can lead to unnatural text \cite{welleck2019neural}, inspiring alternative sampling strategies, \textit{e.g.} top-$k$ or nucleus sampling \cite{holtzman2020curious}. Meanwhile, parallel decode ability of Equation~\ref{eq:prefill-attention} motivates speculative decoding techniques for increasing token throughput \cite{yan2025decoding}.

\hi{Sampling Strategies.}
Greedy sampling selects the index of $\vec{p}$ holding the largest logit, \textit{i.e.} $\argmax_{0\leq i\leq |p|} \vec{p_i}$. But to make the resulting text appear more natural, \textit{(1) probabalistic sampling} stochastically selects the next token according to the actual logits, \textit{i.e.} if $\vec{p_i}=0.1$, then there is a non-negligible chance that $i$ is selected, despite the low logit. This strategy can increase the diversity of the output text but risks sometimes selecting inappropriate tokens\footnote{To offer finer control, a temperature setting can be used to adjust the base logits towards increasing or decreasing the likelihood of selecting low-probability tokens \cite{peeperkorn2024temperature}.}. Alternatively, \textit{(2) top-$k$ sampling} performs probablistic sampling only over the top-$k$ logits, directly preventing low-probability tokens from being selected. Under \textit{(3) nucleus sampling}, the $k$ parameter is set to the minimum number of indices such that their logit sum exceeds a given threshold \cite{holtzman2020curious}.

\begin{figure*}[!t]
\centering
\includegraphics[width=1.0\textwidth]{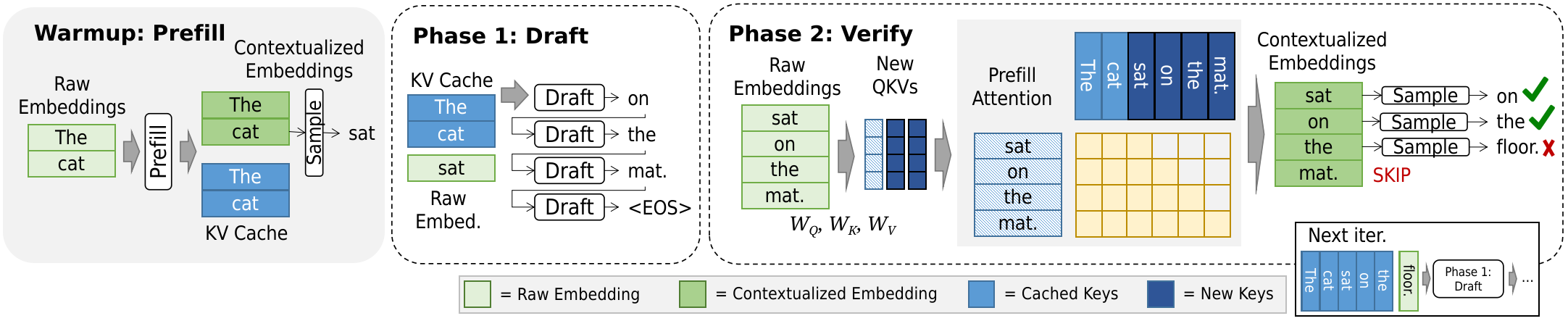}
\caption{Speculative decoding using a draft model followed by parallel verification via Equation~\ref{eq:prefill-attention}.}
\label{fig:speculative-decoding}
\vspace{-1em}
\end{figure*}

\hi{Speculative Decoding.}
The algorithm for speculative decoding is given and analyzed in \cite{leviathan2023fast}. Briefly, given $x_1\dots x_i$:
\begin{enumerate}
\item Approximate the next $b$ tokens, $\hat{x}_{i+1},\dots,\hat{x}_{i+b}$ using a separate process, such as a small language model.
\item Run the main model $b+1$ times in parallel, so that $x_{i+1} = \LLM(x_1\dots x_i)$, $x_{i+2} = \LLM(x_1\dots x_i\hat{x}_{i+1})$, and so on, up to $x_{i+b+1} = \LLM(x_1\dots x_i\hat{x}_{i+1}\dots\hat{x}_{i+b})$.
\item Determine the smallest $j$ such that $\hat{x}_{i+j}\neq x_{i+j}$, then return $x_1\dots x_{i+j}$.
\end{enumerate}
Crucially, the $b+1$ parallel executions in Step 2 produce exactly the set of contextualized embedding vectors, one for each $\hat{x}_{i+j}$, $0\leq j\leq b$, and can be obtained by a single call to $\LLM(x_1\dots x_i\hat{x}_{i+1}\dots\hat{x}_{i+b})$, as during a prefill \cite{yan2025decoding} (Figure~\ref{fig:speculative-decoding}). From these vectors, each of the $x_{i+j}$ tokens can be obtained by sampling.

Techniques for the drafting (Step 1) and verification (Steps 2 and 3) phases are surveyed in \cite{xia2024unlocking,yan2025decoding,zhou2024survey}.

\subsection{Sequence Generation}
\label{sec:generation}

To produce an entire output sequence, the partially generated sequence can be recursively fed back into the model to produce the next token in a streaming manner, stopping once a termination condition is reached, \textit{e.g.} a termination token is produced. But small differences in the prompt, \textit{e.g.} chain-of-thought (CoT) \cite{wei2022chain}, few-shot examples \cite{xu2024does}, and other prompt engineering techniques \cite{sahoo2025systematic}, can produce large differences in the output, with consequences for both cost and accuracy.

On the other hand, structured generation approaches keep track of a set of potential output sequences, autoregressively advancing each candidate sequence as far as needed before selecting one of the candidates to return as the final output sequence. These approaches, \textit{e.g.} beam search \cite{graves2012sequence}, tree-of-thoughts \cite{yao2023tree}, graph-of-thoughts \cite{besta2024graph}, and self-consistency \cite{wang2023self}, multiply the request processing costs but can lead to higher quality outputs for many tasks.

\hi{Streaming.}
In \textit{(1) chain-of-thought} (CoT) prompting, key phrases such as \textit{show your work} are inserted into the prompt in order to elicit longer responses from the LLM, increasing the chance of generating text that better accomplishes the task conveyed by the prompt \cite{wei2022chain}. Likewise, \textit{(2) few-shot examples} appends a number of completed instances of the desired task to the prompt \cite{xu2024does}, with similar aims. But aside from directly manipulating the prompt, the model can be trained to simulate the effects of these techniques without explicitly needing to insert these key phrases. For example, \textit{(3) internalized CoT} is a technique where the model is trained or fine-tuned to generate outputs that resemble chains of thoughts via reward-guided reinforcement learning \cite{fu2024efficiently}. We refer interested readers to \cite{sahoo2025systematic} for a survey of prompt engineering techniques.

\begin{figure}[!t]
\centering
\includegraphics[width=0.48\textwidth]{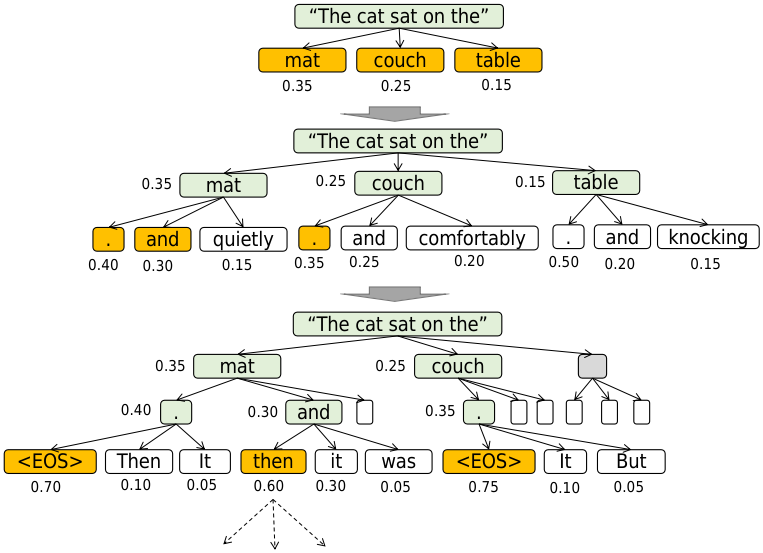}
\caption{Example of beam search with $k=3$.}
\label{fig:beam-search}
\vspace{-1em}
\end{figure}

\hi{Structured.}
Structured approaches construct a pool of candidates from which the final output can be selected. \textit{(1) Beam search} keeps track of $k$ candidate sequences at a time, as shown in Figure~\ref{fig:beam-search}. For each sequence, a number of potential output tokens is sampled, and then the overall top-$k$ tokens out of all the samples are kept, advancing the corresponding sequences while pruning the rest. Once the $k$ sequences have been fully advanced, the sequence with largest cumulative logits is returned as the final output. An algorithm for beam search is given in \cite{graves2012sequence} and variations are given in \cite{freitag2017beam}. \textit{(2) Tree-of-thoughts} \cite{yao2023tree} organizes candidate sequences within a search tree. Each node of the tree represents a potential output, and various prompting strategies are applied to each node to generate lower nodes. \textit{(3) Graph-of-thoughts} \cite{besta2024graph} introduces additional operations over the nodes, such as aggregating from multiple parents, by viewing the nodes as a logical graph. To avoid low-quality candidates, the \textit{(4) self-consistency} technique prunes candidates that are most unlike the others in order to encourage remaining candidates to converge onto a high-quality output \cite{wang2023self}.

\subsection{Discussion}
\label{sec:processing-discussion}

The need to support autoregressive request processing defines LLM inference systems and distinguishes them from other large-scale data processing systems. As a result of this characteristic, the cost of processing a request becomes tied to the output length. But crucially, except for cases where the output sequence is constrained by the application (see Section~\ref{sec:frontends}), the final length cannot be known beforehand. The complicated interactions between the prompt embeddings as a result of the attention operator pose an obstacle to analytically determining the number of rounds before termination. This fundamental uncertainty, especially around memory, motivates techniques like paged attention (Section~\ref{sec:paged-attention}) and job rebalancing (Section~\ref{sec:scheduling}) that try to adapt to changing memory conditions as they arise, in addition to techniques that aim to predict total request memory cost (Section~\ref{sec:execution-discussion}), or generally reduce these costs in the first place (Section~\ref{sec:quantization}).

At the same time, the attention, FFN, and token sampling techniques, along with sequence generation techniques, are self-contained and can be adopted without much difficulty by any inference system. But their unique characteristics must be taken into account during the design of the system. For example, speculative decoding requires a separate drafter model, complicating the execution workflow, and ultimately cannot guarantee higher throughput. Thus, it may be more suited to applications where the drafter has a high chance of success, \textit{e.g.} common-knowledge retrieval-based tasks, instead of tasks that rely on output diversity, \textit{e.g.} creative synthesis. There are similar concerns for techniques that ultimately trade accuracy for efficiency, \textit{e.g.} sparse attention and MoE, as well as for techniques that trade efficiency for accuracy, \textit{e.g.} structured generation.

Looking beyond the current landscape, the continuing spread of LLMs across diverse industries and even beyond natural language \cite{luo2024projecting}, combined with the trend towards specialized hardware \cite{kwon2025lol}, offers increasingly more use cases that could potentially benefit from targeted operator designs and sequence generation strategies, including reinforcement-learning approaches \cite{deepseekai2025deepseek}, novel prompting strategies \cite{zhang2024reactable}, and other approaches. Any new techniques emerging in these areas could lead to a reevaluation of inference system design.

\section{Model Optimization and Execution}
\label{sec:execution}

General purpose CPUs are far outpaced by GPUs and other extreme parallel processor architectures for operations that are primarily data parallel in nature. The desire to take maximum advantage of this capability leads to techniques for kernel design (Section~\ref{sec:kernels}) that aim to develop optimized kernel programs for GPU execution of inference operators. The extreme compute capacity offered by these devices also leads to techniques for request batching (Section~\ref{sec:batching}) that aim to saturate this capacity while paying attention to issues such as stragglers and memory oversubscription due to growing KV caches. Similar considerations inspire techniques for request scheduling (Section~\ref{sec:scheduling}), including techniques for job prioritization and load balancing. These techniques hinge on the ability to accurately predict the total memory cost of a request over its lifetime, likewise the number of execution rounds. In the absence of accurate predictions, mitigation strategies can be used to adopt to new memory conditions as they arise. These techniques are collectively discussed in Section~\ref{sec:execution-discussion}.

\subsection{Kernels}
\label{sec:kernels}

Physical operator cost includes a variety of factors beyond fundamental factors arising from operator design, including running I/O costs, caused by reading and writing intermediate products during operator execution, and invocation costs, caused by loading and unloading intermediate programs in and out of processor cores. For LLM inference, the I/O costs become especially relevant due to the large size of the matrix products (\textit{i.e.} activations) that can result from the attention and FFN operators. For GPU-based workflows, the invocation costs can also become significant due to multiple kernels that may need to be individually launched in order to complete a single operator.

Kernel fusion \cite{boehm2023optimizing} pipelines multiple operations into a single kernel, thus simultaneously avoiding both materialization and invocation costs. This technique, combined with techniques for tiled matrix multiplication \cite{osama2023streamk} and online softmax \cite{milakov2018online}, inspire the development of blockwise attention kernels \cite{hong2024flashdecoding,sanovar2025lean,ye2025flashinfer} that offer vast speedups compared to using generic GPU kernels, as well as distributed attention kernels \cite{liu2023ring} that aim to accelerate inference via multiple GPUs. These techniques have also inspired kernels for other operators beyond attention, including for the FFN \cite{liu2023blockwise} in addition to non-matrix operators \cite{wang2021lightseq,zeng2022boosting,zhai2023bytetransformer}.
\begin{figure}[!t]
\centering
\includegraphics[width=0.45\textwidth]{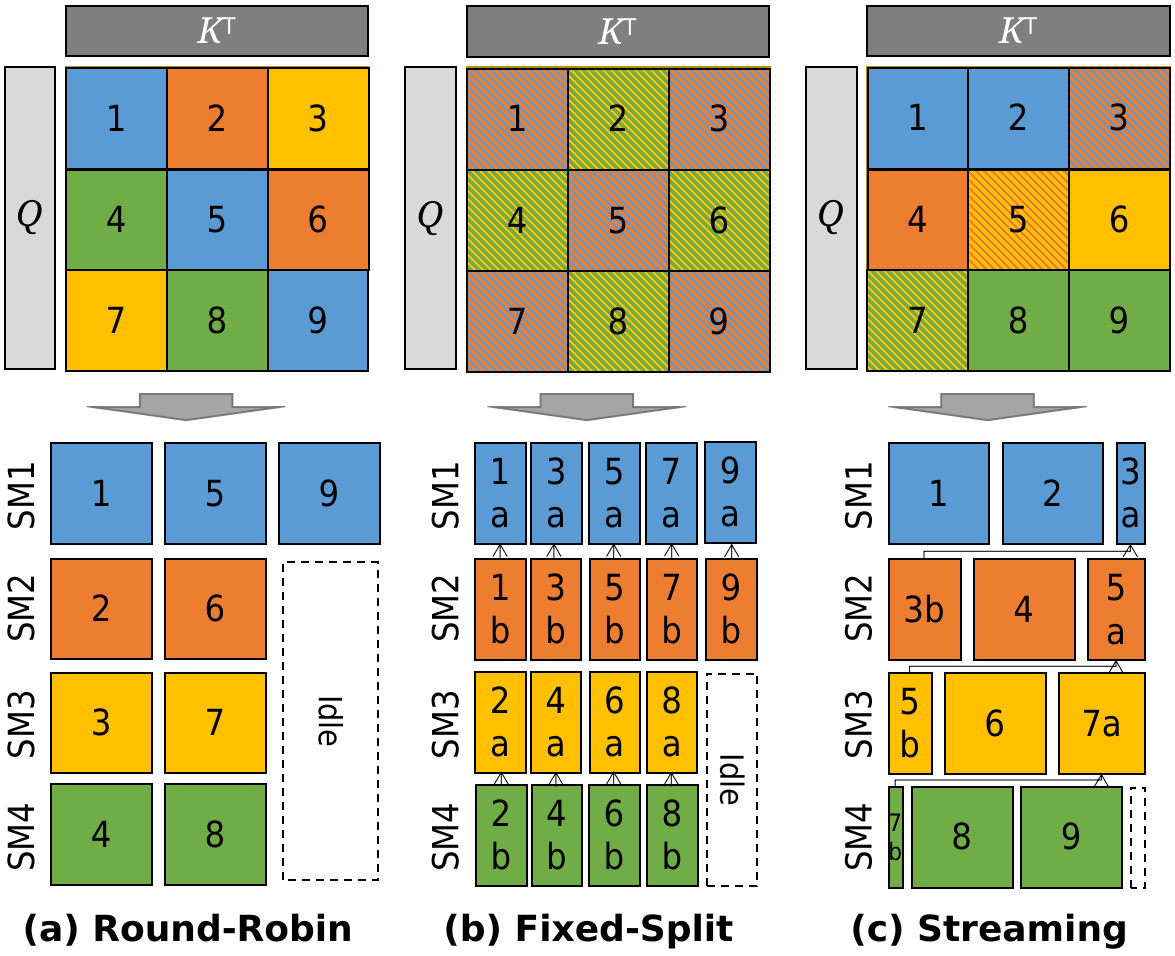}
\caption{Assignment strategies for tiled matrix multiply.}
\label{fig:tiled-matrix-multiply}
\vspace{-1em}
\end{figure}

\subsubsection{Attention Kernels}

\begin{figure*}[!t]
\centering
\includegraphics[width=0.9\textwidth]{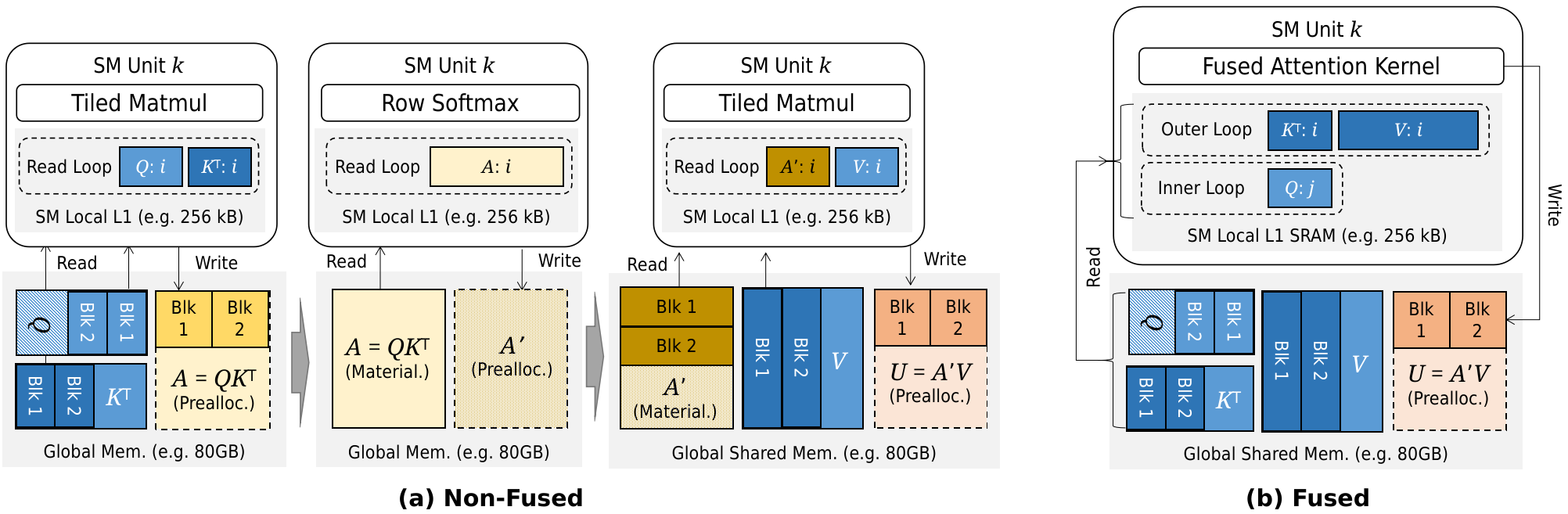}
\caption{Attention operator using non-fused (\textbf{a}) and fused (\textbf{b}) kernels. The fused operator combines tiled matrix multiplication with online softmax to allow writing $U$ in a blockwise fashion, without needing to materialize the intermediate attention pattern.}
\label{fig:fusion}
\vspace{-1em}
\end{figure*}

The attention operator given in Equation~\ref{eq:prefill-attention} requires calculating two matrix products, along with softmax. While the individual scalar products and sums taking place during matrix multiplication are commutative and associative, allowing for trivial parallelization, naively distributing the scalar operations across the processor cores (\textit{i.e.} SM units) of a GPU can lead to low utilization due to stragglers. Moreover, materializing the intermediate matrix products requires costly I/O. Blockwise attention kernels like FlashAttention \cite{dao2022flashattention} fuse online softmax \cite{milakov2018online} with tiled matrix multiplication \cite{osama2023streamk} to address both these concerns. Meanwhile, requests with extremely long contexts (Note~\ref{ft:long-contexts}) may require distributing the KV cache across multiple GPUs. Distributed attention kernels, \textit{e.g.} Ring Attention \cite{liu2023ring}, are designed to handle these situations while dealing with transfer and synchronization costs.

\hi{Blockwise Attention.}
To improve core utilization, Stream-K \cite{osama2023streamk} divides the matrix product into cache-local tiles and sequentially assigns the scalar products in each tile to the processor cores, shown in Figure~\ref{fig:tiled-matrix-multiply}(c). This streaming mechanism is shown to eliminate idle workers that could result from other strategies, such as round-robin assignment (Figure~\ref{fig:tiled-matrix-multiply}(a)) or fixed-split (Figure~\ref{fig:tiled-matrix-multiply}(b)) assignment.
While this technique can be directly used for the matrix products inside the attention mechanism, it still requires materializing the full $QK^\top$ intermediate product before softmax can be applied (Figure~\ref{fig:fusion}(a)). In \cite{milakov2018online}, a numerical approach is given for computing softmax online, where the normalizing terms are collected in an online manner, without the need for full materialization. Online softmax opens up the possibility for tiling the fused attention product (Figure~\ref{fig:fusion}(b)), explored in FlashAttention~\cite{dao2022flashattention} with round-robin assignment, FlashDecoding~\cite{hong2024flashdecoding} with fixed-split assignment, and Lean Attention~\cite{sanovar2025lean} and FlashInfer~\cite{ye2025flashinfer} with streaming assignment.

There are now a number of downstream kernels for other attention variants, mainly sparse attention, that primarily adopt these techniques. For example, FlashMask \cite{wang2024flash} generalize fused attention to support attention masks\footnote{Applying a modifier or mask function $M(\cdot)$ to $QK^\top$.}. Given an arbitrary mask operator, FlexAttention \cite{dong2024flex} compiles a new fused attention kernel according to the operator. The automatically generated kernels offer similar speed-ups compared to handcrafted kernels, like FlashAttention.

\begin{figure}[!t]
\centering
\includegraphics[width=0.48\textwidth]{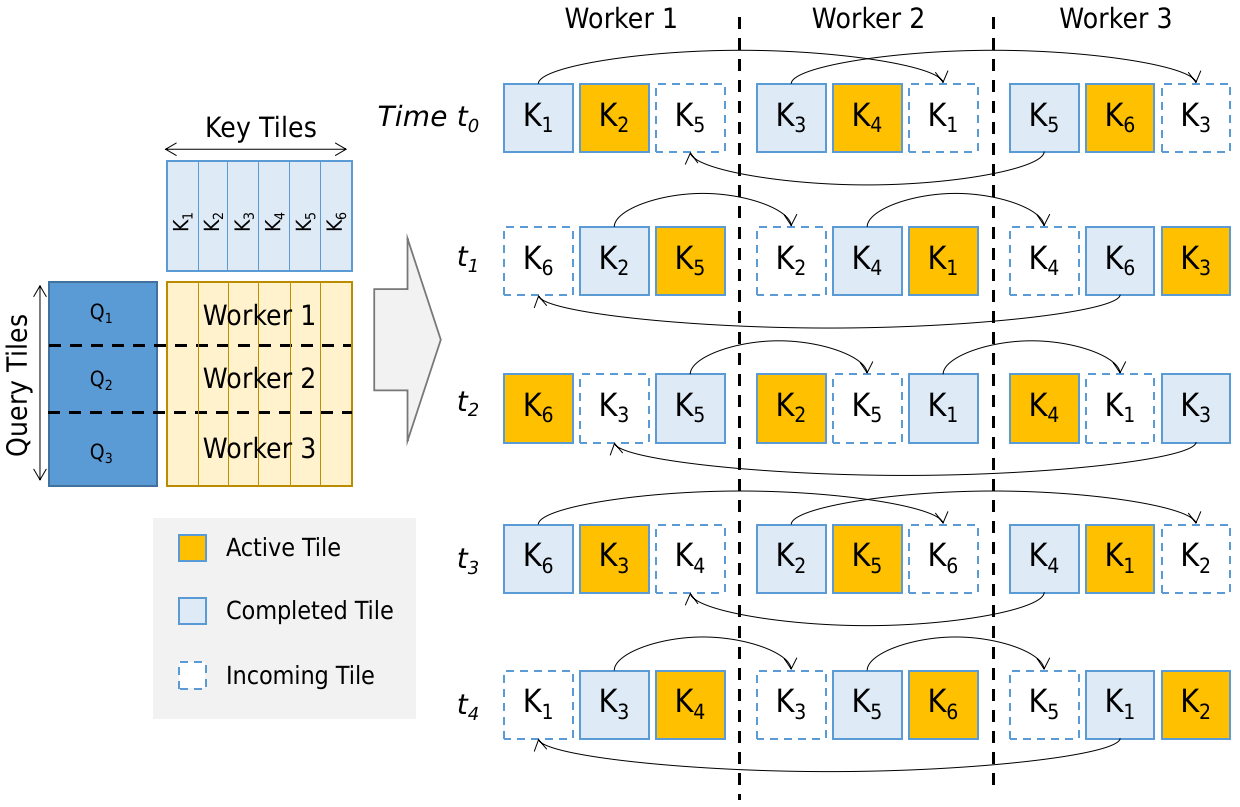}
\caption{Simplified example of ring attention, showing distributed $QK^\top$. At each time step, worker $i$ asynchronously transfers a processed input tile to the next worker in sequence. Eventually, every worker processes every input tile while only requiring enough storage capacity to hold a few tiles at a time.}
\label{fig:ring-attention}
\vspace{-1em}
\end{figure}

\hi{Distributed Attention.}
Ring Attention~\cite{liu2023ring} considers the case where the attention operator is distributed across devices, not just parallelized over processor cores of a single device. The $Q$ matrix is tiled row-wise, so that each worker device gets a subset of queries corresponding to a prompt segment, and then the $K$ matrix is tiled column-wise and distributed to the workers.

To write an attention row, each worker must process every one of the $K$ matrix tiles, requiring an exchange mechanism that can coordinate the transferring of tiles from one worker to the next. Ring Attention adopts a decentralized mechanism where workers transfer tiles in a deterministic sequence while overlapping communication with tile processing in order to reduce effective overhead. Figure~\ref{fig:ring-attention} illustrates a simplified example.

\subsubsection{Other Kernels}

For the FFN, the $f_1$, $f_2$, and $g$ functions can all be implemented element-wise, making it possible to pipeline them into a single fused kernel. This property also allows it to be fused directly to the blockwise attention kernel, leading to modest latency reductions \cite{liu2023blockwise}.

For other operators, LightSeq \cite{wang2021lightseq} combines consecutive non-GeMM operations into single fused kernels, resulting in several handcrafted kernels for operations including layer normalization, tensor reshaping, softmax, and ReLU activation. The result is a reduction in kernel invocations by a factor of 4 per transformer block compared to an implemention based on vendor kernels. Similar techniques are used in \cite{fang2021turbotransformers,zeng2022boosting,zhai2023bytetransformer}.

In addition to fused kernels, DeepSpeed-Inference \cite{aminabadi2022deepspeed} exploits CUDA Graphs\footnote{\url{http://developer.nvidia.com/blog/cuda-graphs}} to launch multiple kernels with a single invocation.

\subsection{Batching}
\label{sec:batching}

\begin{figure*}[!t]
\centering
\includegraphics[width=1.0\textwidth]{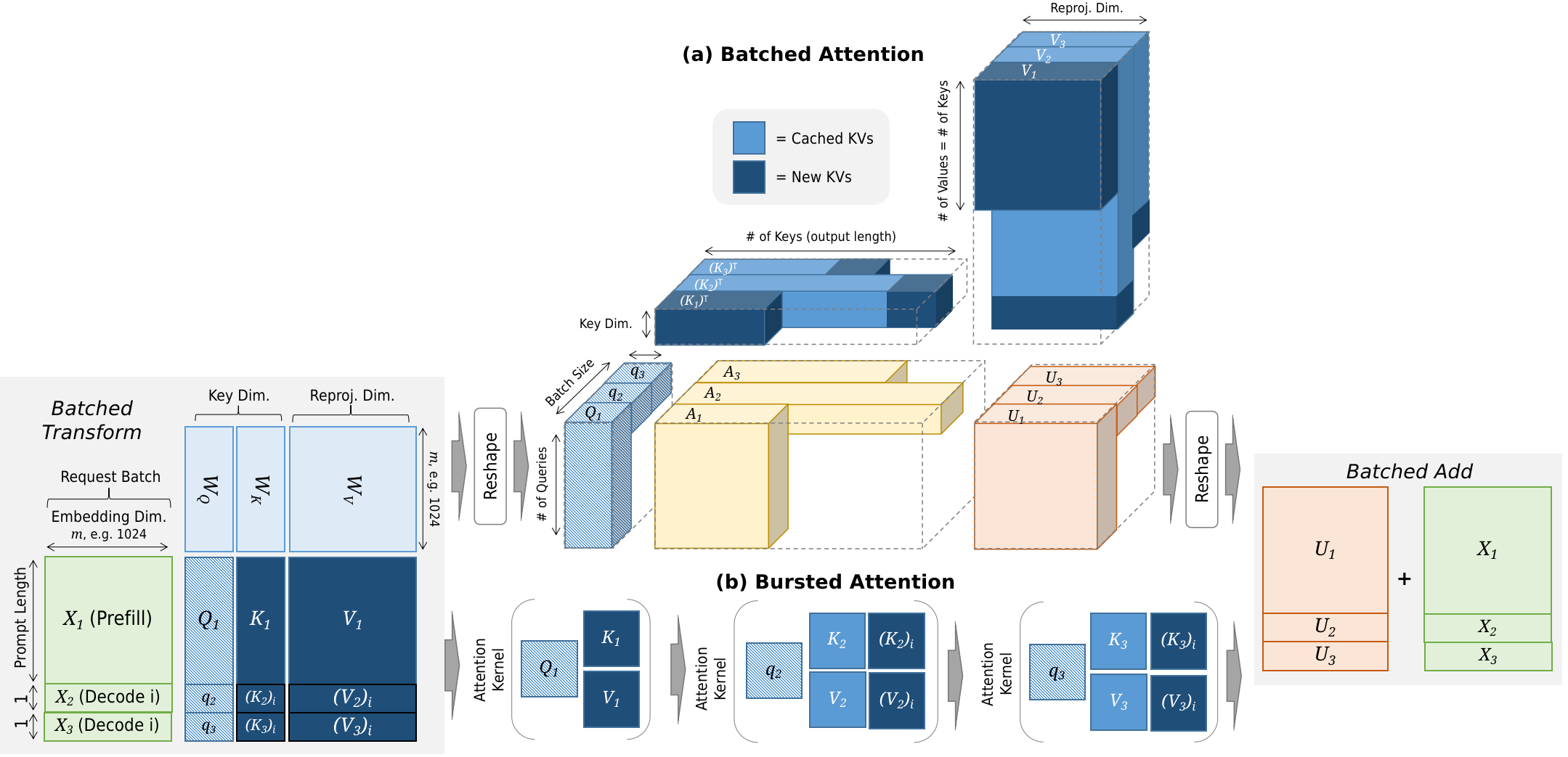}
\caption{Attention over a batch. Request-wise update vectors can be calculated in a single attention kernel (\textbf{a}, showing a non-fused kernel) by concatenating input matrices along the batch dimension, or by using a separate kernel for each request (\textbf{b}).}
\label{fig:batched-transformer}
\vspace{-1em}
\end{figure*}

Before calculating the attention pattern, the input embeddings are multiplied by $W_Q$, $W_K$, and $W_V$ in order to produce the low-dimensional query, key, and value vectors. These vectors can be produced by calculating a single matrix product, and for multiple batched requests, by horizontally concatenating the weight matrices and vertically concatenating the token embeddings. The resulting vectors can likewise be concatenated, along with vectors in the KV cache, to form 3-dimensional matrix inputs for batched attention. Figure~\ref{fig:batched-transformer}(a) illustrates this technique.

If the input to the batched attention operator is ragged, in other words the lengths or widths of the query and key matrices per request are not equal, then calculating the matrix product can lead to underutilized processor cores due to matrix sparsity. This situation arises frequently as a result of unequal request KV cache sizes for requests in the decoding phase as well as unequal prompt lengths for requests in the prefill phase. On the other hand, bursting the batch avoids sparse matrices but requires separate kernel launches, one for each request in the batch \cite{yu2022orca} (Figure~\ref{fig:batched-transformer}(b)).

Additionally, batching frequency and sizing can affect issues like memory oversubscription and latency due to stragglers. Under static batching, every request in a batch is executed until completion before the next batch is processed, leading to the possibility of stragglers delaying a batch. Moreover, since the number of decoding rounds is unknown, static batching risks oversubscribing the memory container as memory usage rises after each round. Dynamic batching \cite{agrawal2024taming,yu2022orca} can be used to mitigate the effects of stragglers, while carefully controlling batch size can be used to avoid oversubscription, especially when combined
with job prioritization (Section~\ref{sec:scheduling}) and memory management techniques (Section~\ref{sec:memory}).

\begin{figure}[!t]
\centering
\includegraphics[width=.45\textwidth]{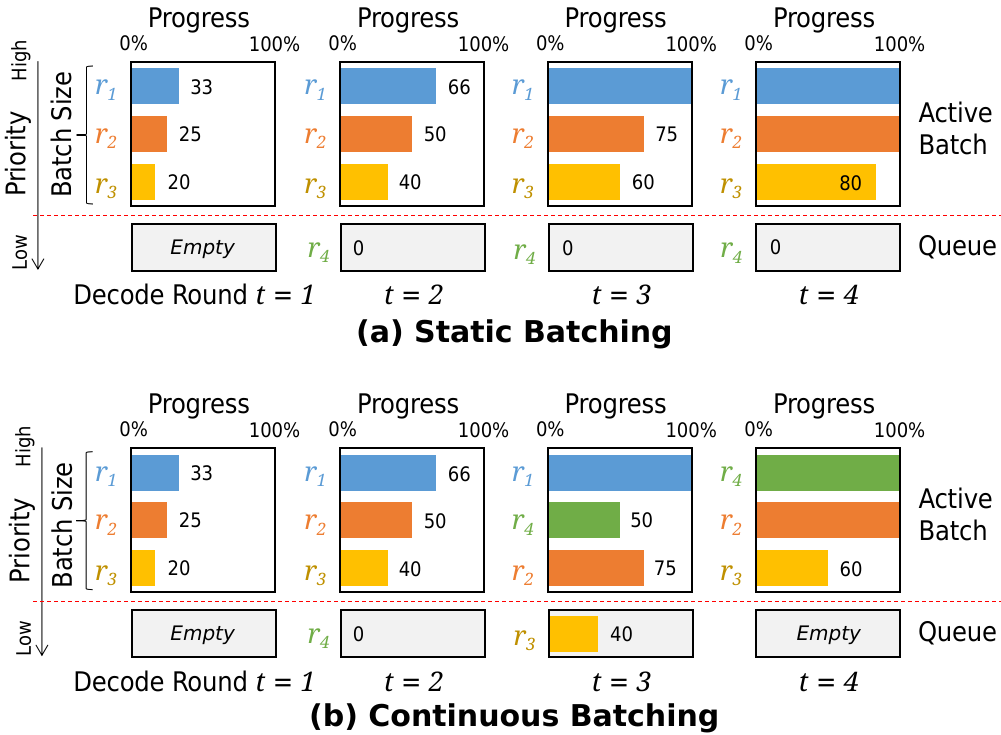}
\caption{Static batching (\textbf{a}) reformulates the active batch once all the requests in the batch are completed, whereas continuous batching (\textbf{b}) reformulates the batch after every decoding round. Note that this figure shows requests prioritized by shortest remaining time.}
\label{fig:batching}
\vspace{-1em}
\end{figure}

\hi{Dynamic Batching.}
With static batching, the requests in the batch are executed and returned together so that their latencies appear equal, even though some requests may have reached termination well before others. To avoid stragglers delaying batch completion, \textit{(1) continuous batching} reconstitutes the batch after each execution round instead of at completion time \cite{yu2022orca} (Figure~\ref{fig:batching}). This simple mechanism offers greater control over request execution. For example, if a request in the batch is completed following an execution round, its KV cache can be immediately evicted, making room for the next queued request to join the batch\footnote{Note that even if a request in the active batch requires additional rounds, the batch controller may still decide to drop the request from the batch, \textit{i.e.} preempt the request, in order to make room for a higher priority request or to reclaim memory. See Figure~\ref{fig:batching}(b), case $t=3$.}. \textit{(2) Chunked prefills} extends the idea of continuous batching to the prefill stage by splitting the prompt into small chunks, processed across multiple execution rounds \cite{agrawal2024taming}.

\hi{Batch Size.}
For static batching, the batch can be formed via greedy bin-packing \cite{jin2023s3}. The batch size is bounded by available memory, as the maximum total memory usage of all the KV caches and intermediate products across the batched requests, for all decoding rounds, must be able to fit inside the memory container.

For dynamic batching, a batch can execute even if the total memory will eventually exceed the memory container because the batch is reconsituted after every round, hence an extremely large batch size is possible by batching many requests with small KV caches. Even so, large batches raises the risk of preemptions due to cache growth, but on the other hand, small batches reduce throughput and can lead to underutilized resources. Most inference systems balance these two risks by fixing the batch size to a constant discovered through offline testing, \textit{e.g.} Orca \cite{yu2022orca}, or alternatively via a fixed token budget, \textit{e.g.} Sarathi-Serve \cite{agrawal2024taming}.

\subsection{Scheduling}
\label{sec:scheduling}

When the rate of requests exceeds system throughput, new requests must wait in a queue before being processed. Since time spent waiting in the queue adds to request latency, the order in which requests are processed affects the latency by shortening or lengthening the waiting times. Job scheduling problems of this nature appear in many domains, and classical techniques, such as first-come first-serve (FCFS), shortest-job first (SJF), and multi-level queuing (MLQ), can be readily adopted for LLM inference \cite{kwon2023efficient,shahout2024dont,wu2024fast}.

But due to the extreme and unpredictable memory pressures caused by KV cache growth, LLM inference systems may face high preemption risk. At the same time, request resumption following preemption can be expensive, requiring either a compute-intensive prefill phase or a bandwidth-intensive transfer from secondary or remote storage in order to recover the lost cache. These factors lead to dynamic promotion mechanisms \cite{fu2024efficiently,kossmann2025gpu,shahout2024dont,wu2024fast} aimed at preventing costly preemptions, in addition to recovery mechanisms (Section~\ref{sec:eviction}).

Additionally, systems equipped with multiple inference replicas (Section~\ref{sec:runtimes}) need load balancing mechanisms for conducting request assignment, for example by greedy least-load assignment \cite{kossmann2025gpu,liu2025mell}. To compensate for inaccurate load prediction, dynamic rebalancing techniques can be used to migrate requests in response to changing worker conditions \cite{liu2025mell,sun2024llumnix}. For systems that support cache sharing (Section~\ref{sec:cache-sharing}), cache availability can also be used to inform load balancing decisions, as maximizing these opportunities helps to reduce overall load \cite{cao2025locality,qin2024mooncake,srivatsa2024preble,zheng2024sglang}.

\begin{figure}[!t]
\centering
\includegraphics[width=0.40\textwidth]{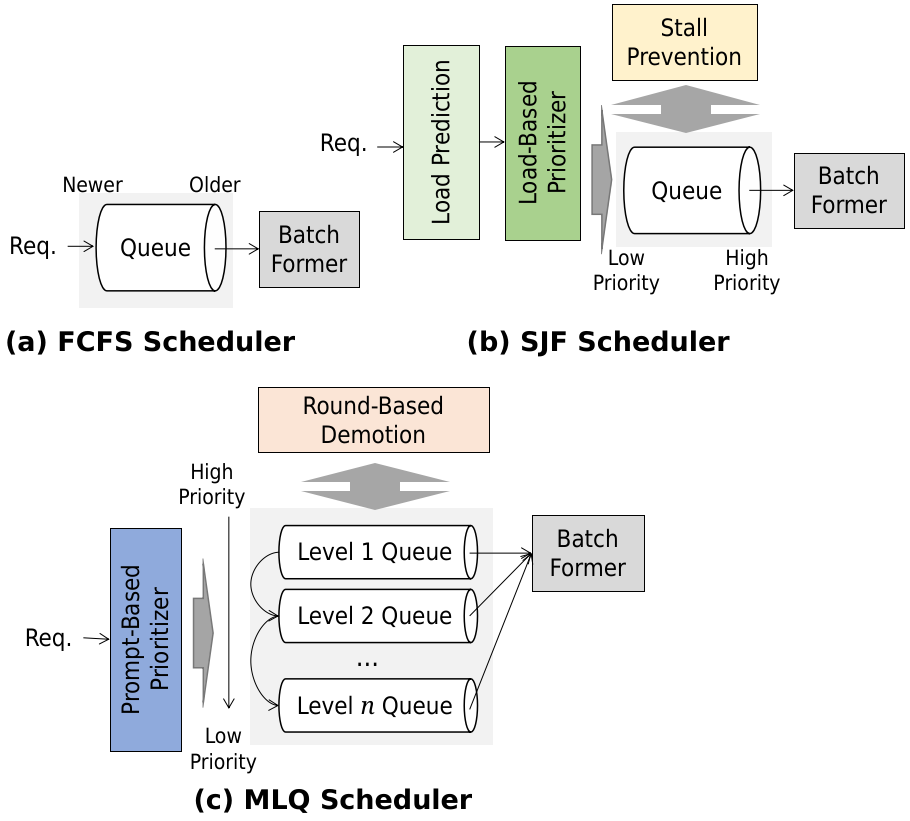}
\caption{Scheduling with FCFS (\textbf{a}), SJF (\textbf{b}), and MLQ (\textbf{c}).}
\label{fig:schedulers}
\vspace{-1em}
\end{figure}

\hi{Job Prioritization.}
Request priorities establish the request processing order while also affecting memory-related preemption, since the lowest priority request is often the one that gets preempted. Request preemption serves to reduce memory pressure by removing the cache entries corresponding to the preempted request. But upon resumption, the cache entries need to be recovered, and whether or not to restore the cache by recomputation or by retrieving from an offloaded container depends on cache size and other factors, as discussed in \cite{kwon2023efficient}.

Under \textit{(1) FCFS}, requests that have spent more time in the queue are given higher priority, but this policy may cause head-of-line blocking as early long-running requests may delay execution of later short-running requests that would otherwise finish quickly. An \textit{(2) SJF} policy avoids this problem and achieves minimal average latency, but requires accurate predictions regarding the number of decoding rounds. We collectively discuss load prediction in  Section~\ref{sec:execution-discussion}. But additionally, SJF also risks job starvation, as requests with high predicted rounds can be continually demoted due to arrival of requests with low predicted rounds. To avoid starvation, the number of times a request can be demoted can be physically limited \cite{shahout2024dont}, or alternatively, long waiting time \cite{fu2024efficiently,kossmann2025gpu,wu2024fast} or large resumption cost \cite{shahout2024dont} can be used as promotion indicators to artificially increase the priority of otherwise perpetually low-priority requests. On the other hand, if accurate job completion times cannot be obtained, an \textit{(3) MLQ} policy can be used to simulate SJF by performing gradual demotion of long-running requests instead of relying on a fixed priority. In \cite{wu2024fast}, this technique is applied to LLM inference by assigning new requests to an initial priority based on prompt length, and then gradually reducing the priority of each active request based on number of produced output tokens.

Aside from these classical scheduling techniques, the high cost of calculating KV vectors motivates \textit{(4) cache-based} policies for systems that support cache sharing which prioritize requests in order to maximize cache hits, thereby avoiding cache thrashing \cite{liu2024optimizing,sun2025hygen,zheng2024sglang}.

\hi{Load Balancing.}
Optimal load balancing is achieved when the maximum load, measured in terms of total compute cost, across the workers is a minimum. When the loads caused by assigning various jobs are known, greedily assigning jobs as they arrive to the least-load worker yields a maximum load (\textit{i.e.} makespan) that is at most twice that of the optimal solution in the worst case\footnote{The factor can be reduced to $4/3$ if jobs are first sorted in descending order by load \cite{graham1969bounds}, but doing so conflicts with SJF scheduling policies that aim to reduce request latency.}. The reasonable performance of greedy load balancing motivates its adoption in multi-replica inference systems but also leads to the unavoidable need for load prediction. Complicating the prediction is the dynamic load changes that occur due to growing KV caches over time, combined with memory fluctuations arising from request preemptions, resumptions, and natural terminations. Since load prediction is important for scheduling as well as load balancing, we collectively discuss these techniques in Section~\ref{sec:execution-discussion}. Outside of reducing makespan, requests can also be assigned to maximize cache hits in order to take advantage of cache sharing \cite{cao2025locality,qin2024mooncake,srivatsa2024preble,zheng2024sglang}.

Requests can also be periodically rebalanced to compensate for inaccurate load predictions. In \cite{liu2025mell}, requests in the decode phase are rebalanced based on changes in their KV cache size. To better characterize worker load, workers are given cache size limits, so that workers with high size limits can be assigned few long-running requests in order to avoid memory-related preemptions while workers with low size limits can be assigned a large number of short-running requests due to the reduced risk preemptions. Initially, requests are assigned to workers based on prefix length. But as cache sizes grow as a result of decoding rounds, requests with caches that cross certain size thresholds are rebalanced to more appropriate workers. In \cite{sun2024llumnix}, requests are periodically rebalanced by forming worker pairs based on load, then migrating requests from the high-load worker in the pair to the low-load worker in the pair until the loads between the two workers are balanced.

\subsection{Discussion}
\label{sec:execution-discussion}

Hardware accelerators have been thoroughly exploited to develop efficient kernels for speeding up LLM inference, but load prediction remains a pressing challenge. Related is the problem of optimal batch sizing for dynamic batching scenarios. Large batch sizes increase throughput but also increase the risk of memory-related preemptions while small batch sizes can lead to underutilized computational capacity. Existing systems tend to establish a token budget, dynamically sizing a batch up to the budget. But this approach is directed more towards achieving balanced TTFT and TBT offered by dynamic batching and less towards balancing throughput with compute capacity. For this latter aim, existing systems resort to offline testing in order to set the budget, \textit{e.g.} Sarathi-Serve \cite{agrawal2024taming}.

Regarding load prediction, one approach is to train a prediction model. To achieve high prediction accuracy, \cite{fu2024efficient} trains a small OPT-125M model \cite{zhang2022opt} to predict the relative ranking of a given prompt with respect to its output length instead of the exact length, and this approach is shown to lead to more robust scheduling compared to using length prediction. A similar approach is used in \cite{jin2023s3} with DistilBERT \cite{sanh2020distilbert}, but the model is trained to select the numerical range that contains the output length instead of predicting a prompt ranking. Instead of using a language model, \cite{shahout2024dont} trains a dedicated MLP using the activations from the LLM as features. In \cite{zheng2023response}, the LLM itself is used for length prediction by appending special prefixes to the prompt\footnote{\textit{E.g.} ``Before responding to the above instruction, you have to predict the length of your response. Print the estimated number of words in your response in the first line.''}. Given the difficulty of load prediction, we note that inference systems that target certain applications can be co-designed with a frontend along with the execution runtime to constrain the outputs so that their lengths can be known with high accuracy. We elaborate on this strategy in Section~\ref{sec:runtimes}.

These types of load predictions can be directly used for SJF scheduling, but in order to use them for load balancing, they should include other factors that account for load fluctuations on the worker. For example, \cite{kossmann2025gpu} combines request memory usage predictions with estimated memory reclamation rates in order to provide a more aggressive measure of memory availability. Mooncake \cite{qin2024mooncake} uses a similar approach while also factoring in time lost due to KV cache transfers.

\section{Memory Management}
\label{sec:memory}

Failing to account for KV cache memory can lead to over-subscription of the memory container, resulting in costly memory swapping. As a preventive measure, the maximum memory of each request can be preallocated at request time, making it impossible to over-subscribe the container \cite{wang2021lightseq}. However, with the exception of length-constrained generation (see Section~\ref{sec:frontends}), the exact final size of the KV cache is unknown, risking memory waste due to over-allocation. Even if it were known, preallocating memory upfront prevents it from being used in the meantime by other requests.

Instead of static preallocation, dynamic paged-based memory management allocates memory in small blocks as the need arises, avoiding low actual utilization caused by large amounts of reserved memory (Section~\ref{sec:paged-attention}). Meanwhile to reduce the memory burden, eviction and offloading techniques (Section~\ref{sec:eviction}) can be used to remove unneeded KV entries from the memory container while quantization techniques (Section~\ref{sec:quantization}) can be used to reduce the physical byte size. In some cases, such as when there is a shared system prompt or in certain RAG workflows, cache persistence techniques (Section~\ref{sec:cache-sharing}) can be used to allow reusing KV entries across multiple requests, avoiding costly recomputation and storage redundancy. The various combinations of offloading, recovery, quantization, and persistence techniques opens up the potential for novel memory schemes that we discuss in Section~\ref{sec:memory-discussion}.

\subsection{Paged-Based Memory Allocation}
\label{sec:paged-attention}

Since the KV cache grows gradually with each decoding round, allocating memory in small blocks as needs arise avoids a large upfront memory cost compared to static preallocation. However, dynamic blockwise allocation can lead to non-contiguous KV caches, requiring a memory manager that can map logical cache entries (\textit{i.e.} cache pages) to physical blocks, as well as paged-based kernels that are specially designed for non-contiguous memory ranges \cite{kwon2023efficient,prabhu2025vattention}. But in addition to enabling dynamic allocation, paged-based memory also opens up opportunities for block sharing \cite{kwon2023efficient} in order to avoid recomputation and storage redundancy. Block sharing forms the basis for cache persistence techniques that will be discussed in Section~\ref{sec:cache-sharing}.

\begin{figure}[!t]
\centering
\includegraphics[width=0.48\textwidth]{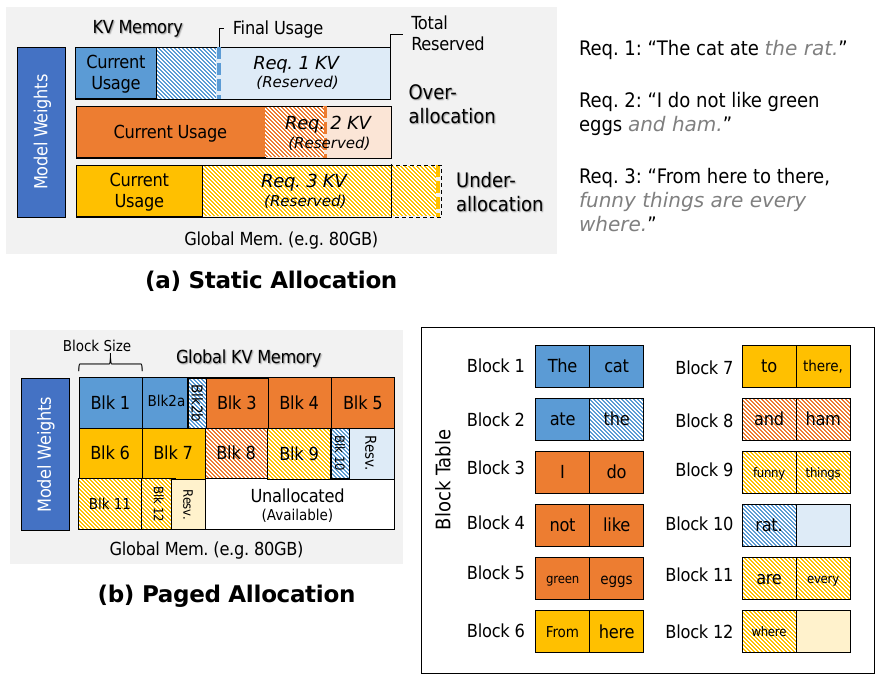}
\caption{Static (\textbf{a}) and paged (\textbf{b}) memory allocation.}
\label{fig:paged-memory}
\vspace{-1em}
\end{figure}

\hi{Memory Manager.}
The memory manager is responsible for page creation, deletion, and lookup. To keep track of page block addresses and contents (\textit{i.e.} tokens that map onto the KV entries stored in the page, along with their positions in the sequence), a page table can be used to list the addresses and contents of each page (Figure~\ref{fig:paged-memory}). In vLLM \cite{kwon2023efficient}, GPU memory serves as the memory container, and the memory manager and page table are implemented on the host machine.

For GPU-based systems using a CPU-based memory manager, since the physical blocks reside on the GPU, page creation and deletion cause memory allocation and deallocation commands to be submitted from the CPU to the GPU. Likewise, page lookup, performed by a special page-aware attention kernel running inside the GPU, causes a lookup command to be submitted from the kernel to the memory manager. In vAttention \cite{prabhu2025vattention}, the  overhead resulting from these communications is reduced by exploiting the native memory management capability of GPUs. Doing so has a secondary advantage of making caches appear to a kernel as if stored in contiguous memory, enabling the use of non-paged kernels like those discussed in Section~\ref{sec:operators}.

\hi{Block Sharing.}
To implement block sharing, multiple page entries corresponding to different requests can simply be assigned to the same block address. Conditions for when to link pages in this manner depend on the shared attention mechanism (Section~\ref{sec:cache-sharing}). Under \textit{(1) exact-match sharing} \cite{kwon2023efficient,zheng2024sglang,srivatsa2024preble}, only the cache entries of the longest common prefix between one or more requests can be shared. Block sizing affects the amount of these sharing opportunities since a page table allows only whole blocks to be shared. Small block sizes make it easier to find blocks where all entries in the block satisfy this condition, but at the same time increase the amount of block retrievals during inference. Under \textit{(2) partial-match sharing}, this condition is relaxed to allow partial prefix matches, \textit{i.e.} prefixes that share some or all of the same tokens that could also be out of sequence.

\subsection{Eviction and Offloading}
\label{sec:eviction}

\begin{figure*}[!t]
\centering
\includegraphics[width=0.9\textwidth]{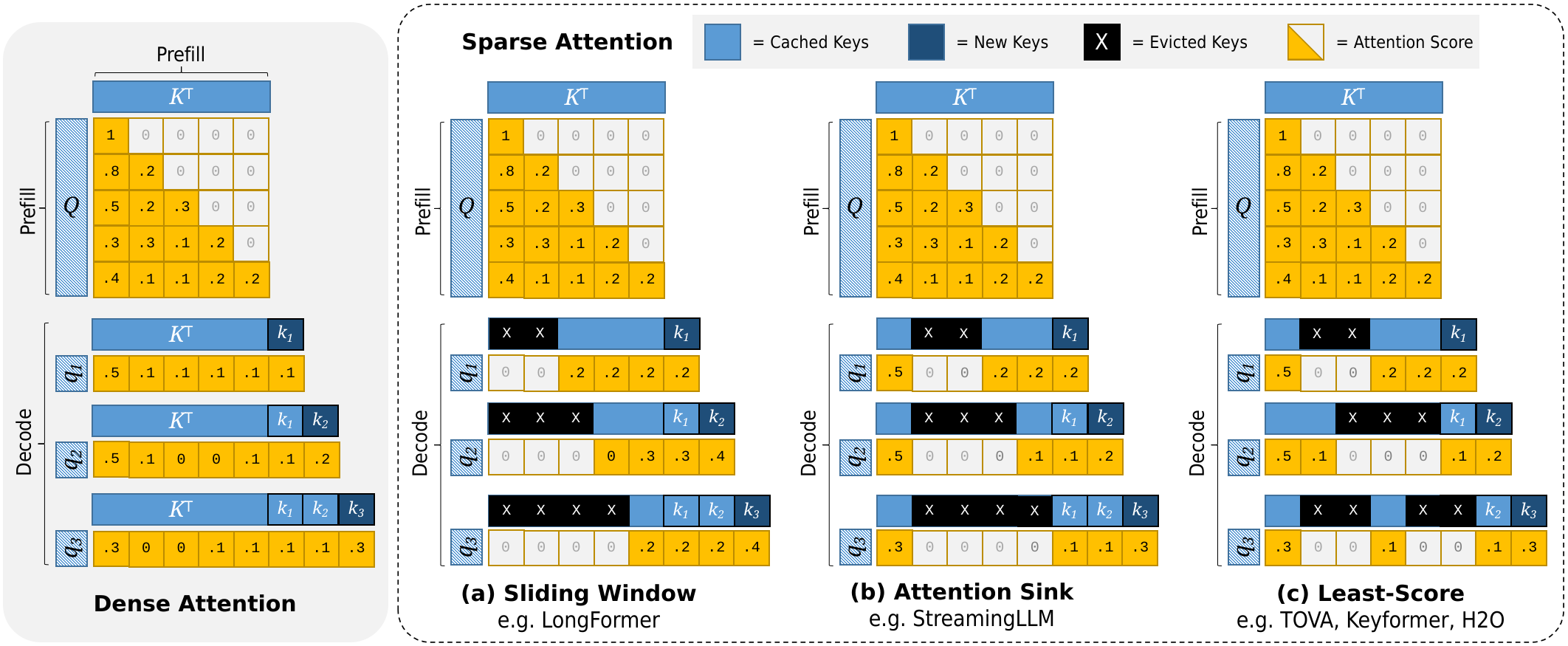}
\caption{Examples of sparse attention using static (\textbf{a}, \textbf{b}) and dynamic (\textbf{c}) masks.}
\label{fig:attention-masks}
\vspace{-1em}
\end{figure*}

Cache eviction and offloading are necessary for supporting preemption, but also useful for supporting long contexts that would otherwise exceed the memory container. In the case of preemption, cache entries of low priority requests can be evicted or offloaded in order to make room for higher priority requests. Since the preempted request must be resumed eventually, whether to evict or offload depends mainly on the recovery cost \cite{kwon2023efficient}. In the case of long contexts, unimportant entries that have little impact on the final output sequence can be evicted in order to allow decoding to continue, \textit{i.e.} sparse attention. Which entries to evict can depend on a variety of factors, including token position \cite{xiao2024efficient} and attention score \cite{oren2024transformers,liu2023scissorhands,zhang2023h2o}. Alternatively, large caches can also be partially offloaded, taking advantage of tiered storage to supplement the memory container. But as with preempted requests, the offloaded entries must eventually be reloaded into the memory container, leading to techniques that aim to balance memory usage with recovery costs \cite{aminabadi2022deepspeed,lee2024infinigen,sheng2023flexgen}. For persisted caches, cache eviction can be used to control the size of the cache, and traditional cache control techniques, \textit{e.g.} LRU, can be adopted for this purpose \cite{zheng2024sglang}.

Evicted entries can be recovered by recomputing the key and value vectors based on the partial LLM output while offloaded entries can be recovered by transferring them back into the memory container. For preempted requests with short prompts that have undergone few decoding rounds, reconstruction may actually be faster than loading from offloaded storage, and this tradeoff is studied in \cite{kwon2023efficient}. For offloaded caches, the effective transfer cost can be reduced by performing the transfer asynchronously ahead of time, overlapping the transfer with other model execution stages in order to hide the transfer latency \cite{lee2024infinigen}. For disaggregated runtimes (see Section~\ref{sec:runtimes}), the decoding worker must recover the cache entries from the prefill worker before decoding can begin. If the decoding worker is selected at the same time as the prefill worker, then the cache can be recovered asynchronously, \textit{i.e.} by streaming entries to the decode worker as they are prefilled \cite{patel2024splitwise}.

\hi{Long Context Eviction.}
Evicting particular cache entries can make room for more significant entries while minimally impacting output quality. \textit{(1) Position-based} policies rank cache entries based on the position of their corresponding tokens with respect to the output sequence. In \cite{xiao2024efficient}, tokens that are near the beginning or trailing end of the output sequence are found to have larger attention values compared to other tokens, leading to an eviction strategy that targets these special positions via handcrafted attention masks (Figure~\ref{fig:attention-masks}(b)). \textit{(2) Attention-based} policies rank cache entries based on their attention values \cite{oren2024transformers} (Figure~\ref{fig:attention-masks}(c)). To increase accuracy, techniques include adding asymmetric noise \cite{adnan2024keyformer}, averaging the score across decoding rounds \cite{ren2024efficacy}, or using accumulative scores \cite{liu2023scissorhands,zhang2023h2o}.

\begin{figure}[!t]
\centering
\includegraphics[width=0.30\textwidth]{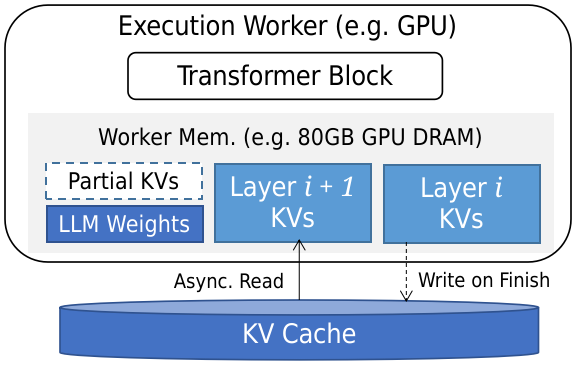}
\caption{Layer-wise KV cache offloading.}
\label{fig:recovery}
\vspace{-1em}
\end{figure}

\hi{Long Context Offloading.}
Alternatively, long context offloading can be used to distribute large KV caches across tiered storage while leaving the cache fully intact. \textit{(1) Layer-wise} offloading moves the cache from some or all transformer layers onto secondary cache storage \cite{aminabadi2022deepspeed,lee2024infinigen}. Since model execution proceeds layer by layer, the cache for layer $i+1$ can be asynchronously transferred during the execution of layer $i$, effectively hiding the transfer latency. Still, the transfers can be costly for large caches, and so \textit{(2) model-wise} offloading only partially offloads the cache but from all layers, allowing transfer costs to be controlled by setting the offloaded fraction \cite{sheng2023flexgen} (Figure~\ref{fig:recovery}).

\subsection{Quantization}
\label{sec:quantization}

\begin{figure}[!t]
\centering
\includegraphics[width=0.48\textwidth]{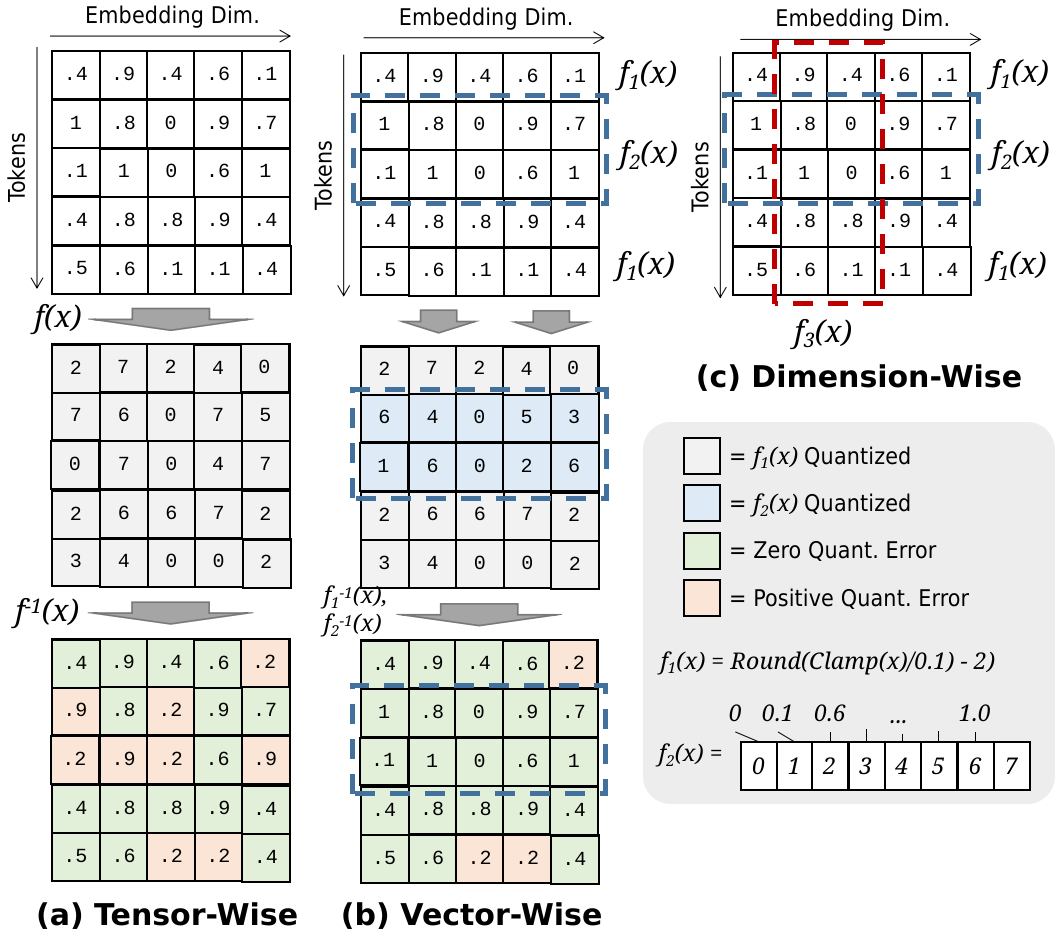}
\caption{Tensor-wise (\textbf{a}), vector-wise (\textbf{b}) and dimension-wise (\textbf{c}) quantization schemes.}
\label{fig:quantization-schemes}
\vspace{-1em}
\end{figure}

\begin{figure}[!t]
\centering
\includegraphics[width=.48\textwidth]{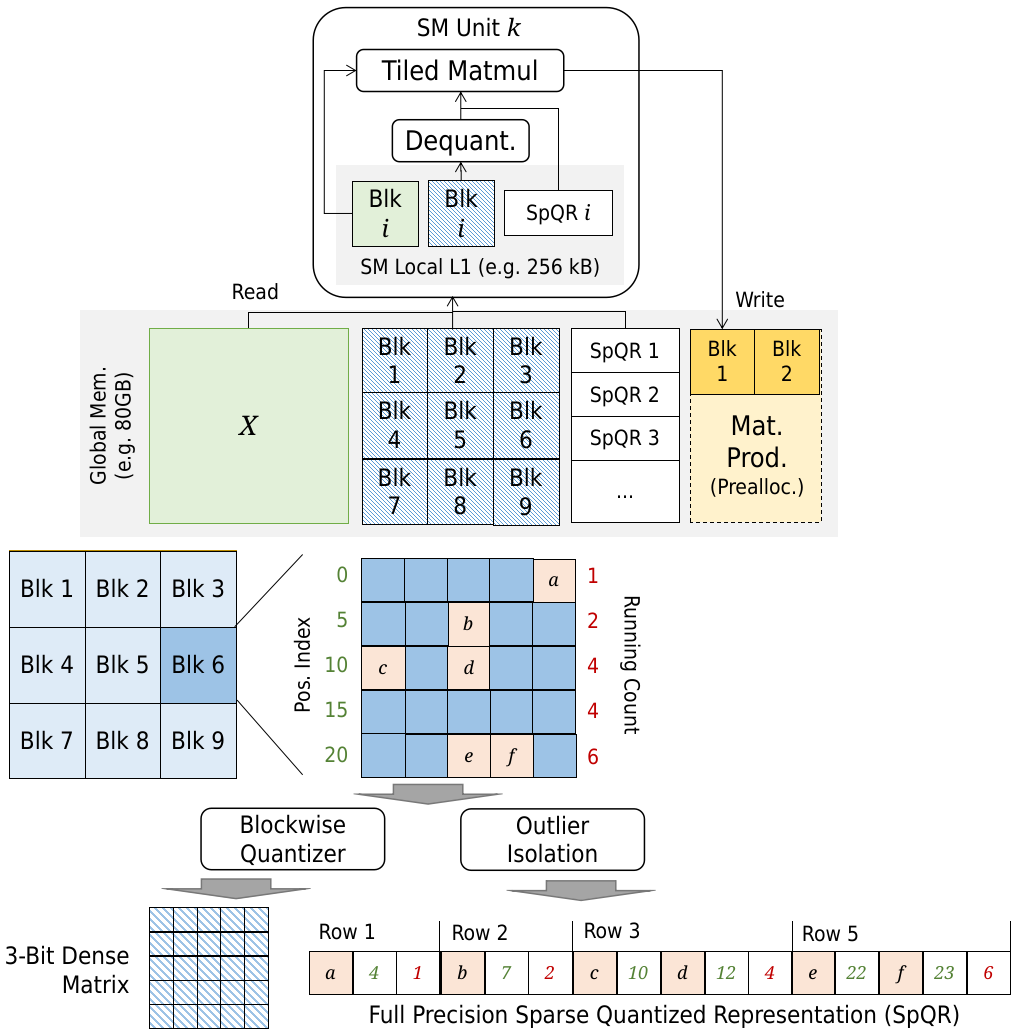}
\caption{Mixed-precision tiled matrix multiply with SpQR.}
\label{fig:mixed-precision}
\vspace{-1em}
\end{figure}

\begin{figure}[!t]
\centering
\includegraphics[width=.48\textwidth]{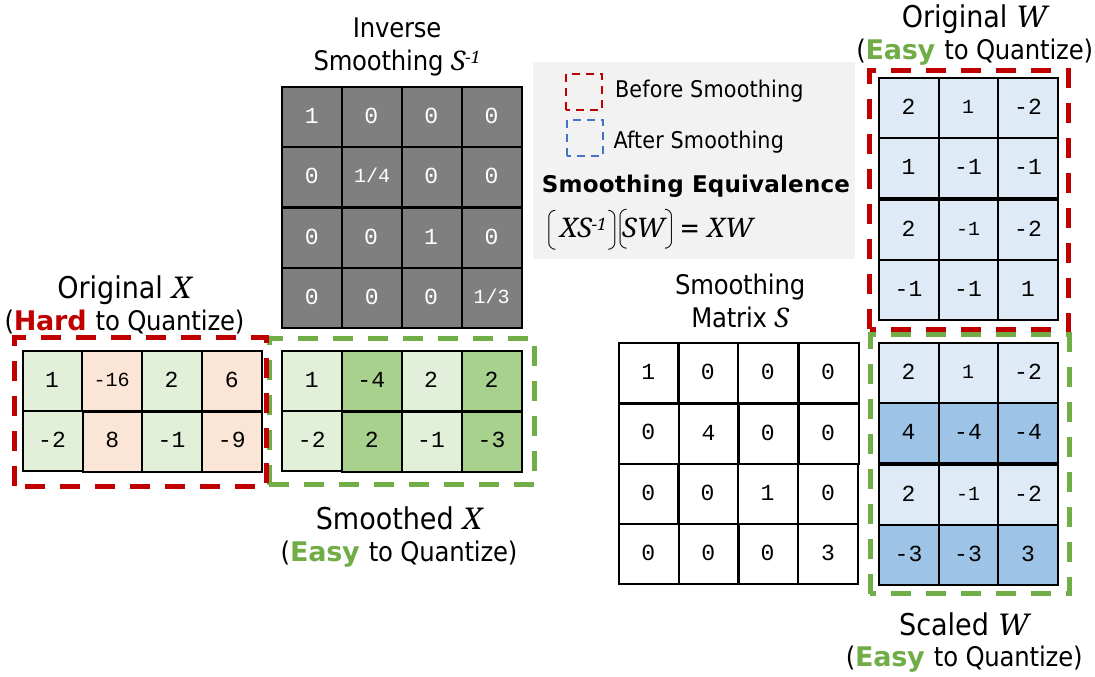}
\caption{Outlier smoothing using a scaling transform.}
\label{fig:outlier-smoothing}
\vspace{-1em}
\end{figure}

Quantization reduces physical byte sizes by lowering the numerical precision\footnote{Models can be directly trained or fine-tuned using quantized weights, e.g. Q8BERT. We refer interested readers to \cite{nagel2021white}.}. To avoid degrading inference quality, quantizers are designed to maximally preserve information-rich regions of the original range within the reduced range. Uniform quantizers map floating-point numbers to a compressed range (\textit{e.g.} integers) via numerical rounding, parameterized by an offset and clamps on the high and low ends of the range \cite{nagel2021white}, whereas non-uniform quantizers aim at further reducing information loss via non-linear methods \cite{frantar2023optq,zadeh2020gobo}. Due to the varying information contents of model components, the quality can also be improved by applying separate quantizers for individual components, leading to quantization schemes based on tensor-wise \cite{nagel2021white,zadeh2020gobo}, vector-wise \cite{dettmers2022llm,yao2022zeroquant}, and dimension-wise quantization \cite{bondarenko2021understanding} (Figure~\ref{fig:quantization-schemes}). Additionally, outlier protection techniques, such as mixed-precision preservation \cite{dettmers2022llm,zadeh2020gobo} (Figure~\ref{fig:mixed-precision}) and outlier smoothing \cite{lin2025qserve,xiao2023smoothquant,zhang2025sageattention} (Figure~\ref{fig:outlier-smoothing}), can be used to prevent quality degradation caused by quantizing significant outlier values.

\hi{Quantizer Design.}
For both uniform and non-uniform quantizers, the design typically aims to minimize a loss function, such as mean-squared error, before and after quantization. For uniform quantizers that take the form $\lfloor x/s \rceil + z$, the small number of parameters makes grid search a viable technique for discovering the optimal values \cite{nagel2021white}. For non-uniform quantizers, techniques such as bucketing \cite{zadeh2020gobo} or more sophisticated search algorithms \cite{frantar2023optq} can be used to directly discover a mapping. Other techniques are covered in detail in \cite{gholami2021survey}.

\hi{Quantization Schemes.}
Since model weights tend to require large amounts of storage, \textit{(1) tensor-wise} quantization over the weight matrices can directly lead to large storage savings \cite{frantar2023optq,nagel2021white,zadeh2020gobo}. For activation matrices, \textit{(2) vector-wise} quantization
offers finer control over the quality of the quantized matrix by stratifying the matrix into $g$ groups, each containing $n/g$ embedding vectors, and then applying a different quantizer to each of the groups \cite{dettmers2022llm,yao2022zeroquant}. The $g$ value can be adjusted based on the necessary granularity. Higher $g$ values offer finer control over the quantized ranges but require training more quantizers and complicate downstream matrix multiplication operations. For even finer control, \textit{(3) dimension-wise} quantization divides each of the vectors into $k$ segments, effectively partitioning the $d$-dimensional vector space into $d/k$-dimensional subspaces, and then applies a separate quantizer across each of the segments \cite{bondarenko2021understanding}.

\hi{Outlier Protection.}
Outlier values have been shown to disproportionately affect model quality \cite{dettmers2022llm,xiao2023smoothquant,zadeh2020gobo}. \textit{(1) Mixed-precision preservation} leaves outliers in raw form but causes tensors to be in a mixed precision state. Specialized data structures can be used to store these tensors, for example by using separate byte regions for low-precision and high-precision values \cite{dettmers2023spqr,zadeh2020gobo} (Figure~\ref{fig:mixed-precision}), but operating over these structures requires specialized decoders or kernel operators that can recognize the regions. On the other hand, \textit{(2) outlier smoothing} can be used to avoid mixed-precision tensors during matrix multiplication while still preserving information from outliers. Given two matrix operands, the technique applies scalar division to reduce the effective strength of outliers in the high-variance matrix while performing reverse scaling in the low-variance matrix, effectively preserving the high-variance information inside the matrix product (Figure~\ref{fig:outlier-smoothing}). In \cite{xiao2023smoothquant}, this technique is applied to activation and weight matrices while in \cite{lin2025qserve,zhang2025sageattention}, this technique is applied to the key and value matrices that form the KV cache.

\subsection{Cache Persistence}
\label{sec:cache-sharing}

The KV vectors in the inner attention layers take on different values depending on the positions of their respective tokens in the generated sequence. As a result, the KV entries of two requests are identical only up to the first position in their sequences where their tokens differ, leading to the exact-match condition for direct cache sharing. But in some cases, KV entries can still be shareable up to a degree. For example in a RAG workflow, two requests may use the same document chunks, even if the prompts are difference.

\begin{figure}[!t]
\centering
\includegraphics[width=0.48\textwidth]{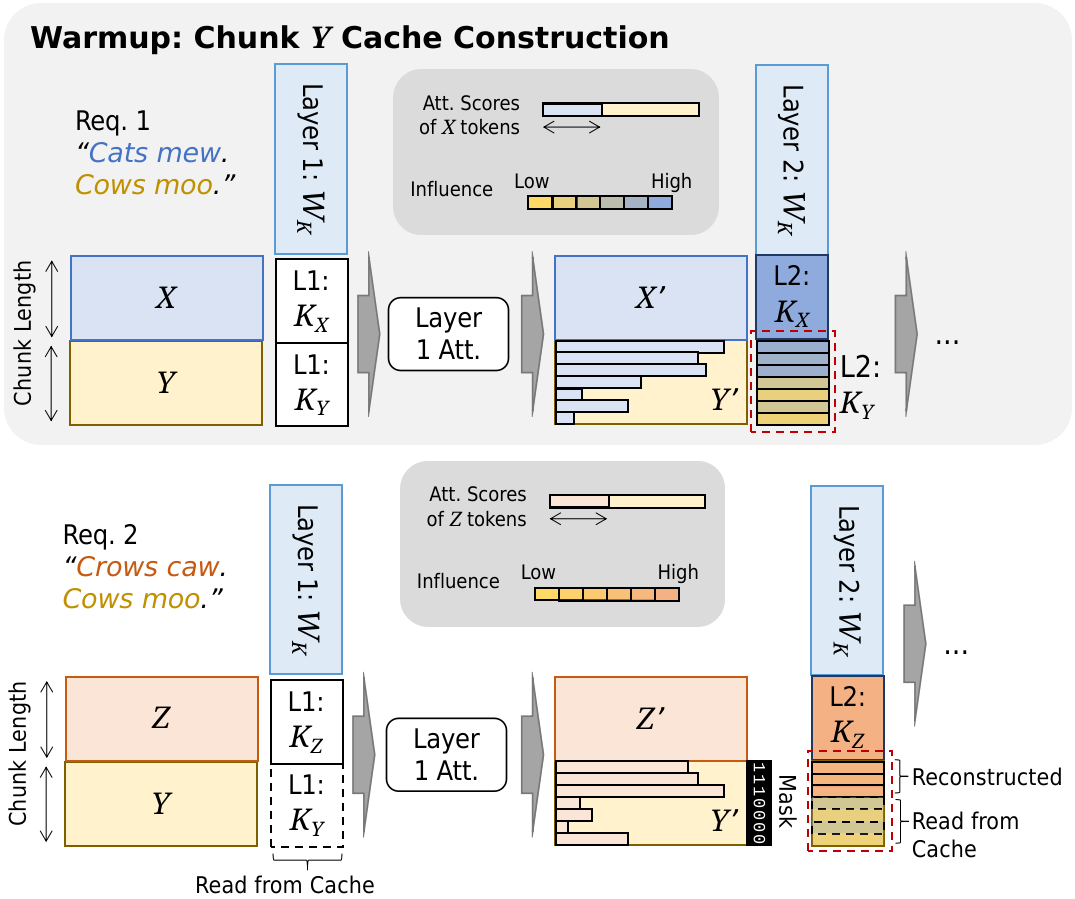}
\caption{The KV entries for chunk $Y$ are influenced by chunk $X$ (top, red box), posing a challenge for cache reuse. Selective entries can be recomputed to mitigate quality degradation (bottom).}
\label{fig:chunk-influences}
\vspace{-1em}
\end{figure}

The basic \textit{(1) prefix sharing} technique reuses persisted cache entries under exact-match prefixes. To find the longest matching prefix, the request prefix can be scanned from beginning to end, performing cache lookups at each token to see if a cache entry already exists for the token at its position in the sequence \cite{liu2024optimizing}. If there are many persisted cache blocks, an index structure, such as a radix tree, can be used to speed up this retrieval \cite{zheng2024sglang}. For partial-match sharing, \textit{(2) selective reconstruction} techniques mitigate quality degradation arising from using non-aligned entries by recalculating KV entries for a subset of most impactful tokens, as shown in Figure~\ref{fig:chunk-influences}. These tokens can be identified by using attention score deviations in the first transformer layer as a ranking criteria \cite{yao2024cacheblend}, or by using token position as a heuristic \cite{gim2024prompt,hu2025epic}.

\subsection{Discussion}
\label{sec:memory-discussion}

The techniques discussed in this section arise out of a need to reduce memory pressure caused by the huge number of model parameters, amounting in the billions for the most common LLMs, in addition to the large sizes of KV caches generated during inference. Paged memory allows for more flexible memory allocation compared to static preallocation, while eviction and offloading, quantization, and cache persistence techniques all aim to directly reduce memory usage.

Moreover, the variety of approaches allows targeting differences in memory capacity, desired accuracy, and context lengths across LLM systems and applications. For preempted requests, since tiered storage is plentiful, offloading can be effective for reducing recovery cost by avoiding cache recomputation, especially since the recovery can be performed asynchronously. For supporting long contexts, eviction and offloading can be appealing alternatives to distributed attention mechanisms, such as Ring Attention \cite{liu2023ring}, that require horizontally scalable GPUs. Quantization can be applied by any inference system, but determining the optimal quantization scheme that achieves a balance between memory savings and model quality may involve trial and error. For example, in \cite{yao2022zeroquant}, weight and activation matrics are both quantized using vector-wise quantization, but as weight values are observed to have less variance compared to activation values, the $g$ term is set to $g>1$ for weights and $g=1$ for activations. In contrast, in \cite{dettmers2022llm}, the $g$ term is set to $g=1$ for both weight and activation matrices to achieve more aggressive space savings, but with outlier protection for activations in order to guard against accuracy loss. The model size and architecture may also affect the quantization scheme. For example, encoder-only BERT models have been shown to be less sensitive to aggressive weight quantization compared to decoder-only GPT models due to lower weight variances. As a result, GPT models seem to benefit from more sophisticated quantization schemes, such as vector-wise or dimension-wise quantization, as demonstrated in \cite{yao2022zeroquant}. Model size has also been shown to be a key consideration when deciding on a particular quantization scheme \cite{dettmers2022llm}. Finally, for cache persistance, if accuracy is not paramount, selective reconstruction can be a useful technique for aggressive memory savings, otherwise prefix matching can be used with no impact on accuracy.

A key decision for inference system design is whether KV cache memory should be constrained or unconstrained. Constraining the cache size makes memory usage more predictable because the maximum memory usage is known, simplifying batching and scheduling decisions, but also potentially reduces model accuracy, especially for long-running requests. For the best of both worlds, we mention an interesting technique. In \cite{cai2024pyramidkv,yang2024pyramidinfer}, the memory constraint is allowed to vary across layers, allowing large memory reductions to be achieved with little impact on output quality simply by allocating more memory to earlier layers than to later ones. Future similar techniques could focus on dynamic memory allocation at the request level, or other allocation schemes. Another approach similar to eviction is entry merging, where instead of discarding evicted entries, they are incorporated or merged into the remaining non-evicted tokens. The goal becomes not to select the least influential tokens, but to select similar tokens, so that the KV entries as a whole are not much different from before the merge, except that there are now fewer entries \cite{liu2024minicache,wan2024d2o,wang2024model,zhang2024cam}.

\section{Inference Systems}
\label{sec:systems}

In an LLM inference system, various request processing, execution, and memory management techniques are combined to support efficient and high-quality processing generic LLM workloads, or to target more specific applications. A complete inference system consists of a frontend and a runtime. The frontend allows for user interaction, \textit{e.g.} through a declarative or imperative interface, and may provide features such as structured outputs and automatic prompt optimization (Section~\ref{sec:frontends}). The runtime manages all the other aspects of model execution, with single-replica runtimes targeted at request processing over a single LLM and multi-replica runtimes targeted at environments with multiple identical LLMs (Section~\ref{sec:runtimes}). Both single and multi-replica systems can support distributed execution, while the latter can also be designed to support disaggregated execution. We summarize the key features of some popular systems in Section~\ref{sec:systems-discussion}.

\subsection{Frontends}
\label{sec:frontends}

\begin{table}[t]
\caption{Frontend features.}
\label{tab:frontends}
\centering
\tabcolsep=0.08cm
\begin{tabular}{llccccc}
\toprule
Frontend
    & Type
    & \shortstack{Ctrl.\\Flow}
    & \shortstack{Struct.\\Out.}
    & \shortstack{Temp.\\Comp.}
    & \shortstack{Prompt\\Opt.}
    & \shortstack{Coupled-\\Run.} \\
\midrule
LMQL \cite{beurerkellner2023lmql}
    & Decl.
    & \cmark
    & \cmark
    & Stag.
    &
    & \\
DSPy \cite{khattab2024dspy}
    & Decl.
    & \cmark
    & \cmark
    & Stream
    & \cmark
    & \\
SGLang \cite{zheng2024sglang}
    & Imp.
    & \cmark
    & \cmark
    & Stag.
    &
    & \cmark \\
Guidance
    & Imp.
    & \cmark
    & \cmark
    & Stream
    &
    & \\
LangChain
    & Imp.
    & \cmark
    & \cmark
    & Stream
    &
    & \\
\bottomrule
\end{tabular}
\end{table}

A frontend allows users to programmatically access the underyling LLM while retaining control over the precise LLM inputs and outputs. Frontends can provide modules or commands that transcribe user instructions into a CoT, few-shot, or RAG prompt, submit different prompts depending on the output of prior prompts, constrain outputs to obey specific templates or formats, and so on. Out of these capabilities, structured outputs and template completion in particular allow for interleaved prompting which can be used to speed up inference. There is a large body of work regarding prompting strategies\footnote{See also \url{http://smith.langchain.com/hub}.} and we refer interested readers to \cite{sahoo2025systematic}.

The interface itself can be imperative or declarative (Table~\ref{tab:frontends}). Imperative interfaces offer lower level APIs that can be used to form an LLM program while declarative interfaces allow for more user-friendly query-based interaction through query strings or high-level modules.

\hi{Basic Interaction.}
Control flow, \textit{e.g.} \texttt{if...else} and \texttt{for} statements, are handled by capturing LLM output and then submitting different prompts depending on the output.
\begin{example}[Capture and Control Flow]
The generated output is captured inside \texttt{s["tool"]} and participates in controlling execution flow\footnote{Example from \url{http://docs.sglang.ai/frontend/frontend}}.
\begin{verbatim}
s += assistant("To answer " + q + ", I need "
    + gen("tool", choices=["calc", "www"]))
if s["tool"] == "calc":
    // .. do something
elif s["tool"] == "www":
    // .. do something
\end{verbatim}
\end{example}

Constrained generation, which stops LLM generation once a user-specified condition is met, is useful for terminating an otherwise lengthy LLM response and can likewise be implemented using output capture. For example, LMQL \cite{beurerkellner2023lmql} exposes this feature via a declarative syntax, \textit{e.g.} the prompt script \texttt{"Give your answer here: [y]" where len(TOKENS(y)) < 25} terminates when the length of the generated output captured by the Python interpreter exceeds 25 tokens.

\hi{Structured Outputs.}
A related feature is structured outputs, where the LLM output is constrained to a user-specified type and format, \textit{e.g.} two-digit number. To implement this feature, the frontend transcribes the constraints into a prompt phrase, \textit{e.g.} ``Give answer as a two-digit number'', then enforces the constraints via logit masking \cite{willard2023efficient}, token resampling, or reprompting, based on the LLM response.

Structured outputs can be extended to support template completion, where the goal is to prompt the LLM to fill in a template, \textit{e.g.} a JSON schema. Since each template item is a structured output, templates can be filled by simply prompting the LLM over each of the items. Alternatively, the LLM can be prompted to fill in the template in a single shot, but this strategy may require multiple reprompting rounds until the LLM generates an output that completely obeys the template, especially since the template itself must appear fully intact within the generated response. Item-by-item prompting also allows for interleaving prompts and decodes in order to speed up inference by taking advantage of fast prefill, especially on runtimes that support persisted caches.
\begin{example}[Constrained Template Generation]
\begin{verbatim}
Write a summary of Bruno Mars, the singer:
{{  "name": "[STRING_VALUE]",
    "age": [INT_VALUE],
    "top_songs": [[
        "[STRING_VALUE]",
        "[STRING_VALUE]" ]] }}
\end{verbatim}
Shown above is an LMQL program\footnote{Example from \url{http://lmql.ai/playground}}. The upper-case bracketed items specify the type constraints. Given this program, LMQL parses it into prompts and coordinates execution as well as output capture. To illustrate the prefill and decode interleaving technique, below is the first prompt submitted by LMQL:
\begin{verbatim}
Write a summary of Bruno Mars, the singer:
{ "name": "
\end{verbatim}
After the LLM outputs \texttt{Bruno Mars}, LMQL submits the second prompt, which is
\begin{Verbatim}[commandchars=\\\{\}]
Write a summary of Bruno Mars, the singer:
\{ "name": "\textcolor{red}{Bruno Mars}",
  "age": "
\end{Verbatim}
where the previous output (shown in red) is directly included in the prompt. Since much of the prompt overlaps with the first prompt, the runtime can take advantage of shared cache entries to speed up prefill. This process repeats until the template is filled.
\end{example}

\hi{Declarative Frontends.}
In a declarative interface, users access these features by calling declarative statements or modules. For example, \textit{(1) LMQL} \cite{beurerkellner2023prompting} uses a declarative syntax, \textit{e.g.} a prompt statement is followed by a \texttt{from} clause which specifies the model and a \texttt{where} clause which constrains the output. These declarative queries can be wrapped inside control flow structures\footnote{See \cite{beurerkellner2023lmql} for examples.}. The LMQL frontend transcribes the query into a prompt chain and enforces the constraints on the generated output. Since the prompt structure is declared upfront, it can be treated as a template, allowing for efficient template generation via prompt and decode interleaving. On the other hand, \textit{(2) DSPy} \cite{khattab2024dspy} offers an object-oriented declarative interface. Different from LMQL, instead of writing prompt queries, the user writes module declarations, with different modules encapsulating different prompt strategies. Prompt content and other parameters are passed into the modules before they are called. This framework allows DSPy to perform prompt optimizations, e.g. synthesizing few-shot examples in order to increase model accuracy.

\hi{Imperative Frontends.}
Users can also access these features through an imperative interface. For example, \textit{(1) SGLang} \cite{zheng2024sglang} provides a low-level API for interacting with the underlying LLM. To improve inference speed, it features a speculative execution technique where the LLM is allowed to continue decoding beyond a user termination condition in the hope of generating additional useful tokens, thereby avoiding multiple API calls. The co-designed SGLang runtime also allows the frontend to take advantage of cache persistence via prompt and decode interleaving for template generation. \textit{(2) Guidance}\footnote{\url{http://github.com/guidance-ai/guidance}} offers similar features as LMQL except it uses an imperative syntax instead of a declarative one. Like LMQL, output constraints are passed as arguments to the generation function. \textit{(3) LangChain} is similar to Guidance but more object-oriented. Prompts and LLM outputs are encapsulated inside classes which can be programmatically manipulated. LangChain also offers an agent-based framework called LangGraph that can be used to program complex interactions, \textit{e.g.} nested \texttt{if/else} ladders.
\subsection{Runtimes}
\label{sec:runtimes}

\begin{table*}[t]
\caption{Model execution features.}
\label{tab:execution-summary}
\centering
\tabcolsep=0.11cm
\begin{tabular}{llllll}
\toprule
Type
  & Runtime
  & Load Bal.
  & Job Prior.
  & Batch Freq.
  & Batch Size \\
\midrule
Single-replica
  & (2022) Orca \cite{yu2022orca}
  & N/A
  & FCFS
  & Cont.
  & Fixed \\
  & (2023) vLLM \cite{kwon2023efficient}
  & N/A
  & FCFS
  & Cont.
  & Fixed \\
  & (2024) Sarathi \cite{agrawal2024taming}
  & N/A
  & FCFS
  & Cont. + CP
  & Token budget \\
  & (2024) SGLang \cite{zheng2024sglang}
  & N/A
  & FCFS
  & Cont. + CP
  & Token budget \\
  & (2024) FastServe \cite{wu2024fast}
  & N/A
  & MLQ (FCFS)
  & Cont.
  & Fixed \\
Multi. mono.
  & (2024) Preble \cite{srivatsa2024preble}
  & Max hit + min load
  & MLQ (Cache Hit)
  & Cont. + CP
  & Token budget \\
Multi. disagg.
  & (2024) DistServe \cite{zhong2024distserve}
  & Min queue (P), min load (D)
  & FCFS (P,D)
  & Cont.
  & Token budget (P,D) \\
  & (2024) TetriInfer \cite{hu2024inference}
  & Min load (P), rand-2 (D)
  & (*) (P), FCFS (D)
  & Cont. + CP
  & Token budget (P,D) \\
  & (2024) SplitWise \cite{patel2024splitwise}
  & Min queue (P + D)
  & FCFS (P,D)
  & Cont. + CP
  & Token budget (P,D) \\
  & (2024) Mooncake \cite{qin2024mooncake}
  & Min TTFT (P + D)
  & FCFS (P,D)
  & Cont. + CP
  & Token budget (P,D) \\
Multi. $\lambda$
  & (2025) DeepFlow \cite{hu2025deepflow}
  & Max hit + min load (P,D)
  & FCFS (P,D)
  & Cont. + CP
  & Token budget (P,D) \\
\bottomrule
\end{tabular}
\end{table*}

\begin{table*}[t]
\caption{Memory management features.}
\label{tab:memory-summary}
\centering
\tabcolsep=0.11cm
\begin{tabular}{llllllll}
\toprule
Type
    & Runtime
    & Mem. Model
    & Mem. Repl.
    & Evict/Offload
    & Recov.
    & Persis. Model \\
\midrule
Single-replica
    & (2022) Orca \cite{yu2022orca}
    & Static (GPU)
    & N/A
    & N/A
    & N/A
    & N/A \\
    & (2023) vLLM \cite{kwon2023efficient}
    & Paged (Tiered)
    & N/A
    & Cost eq.
    & Sync (CPU to GPU)
    & N/A \\
    & (2024) Sarathi \cite{agrawal2024taming}
    & Paged (Tiered)
    & N/A
    & Cost eq.
    & Sync (CPU to GPU)
    & N/A \\
    & (2024) SGLang \cite{zheng2024sglang}
    & Paged (GPU)
    & N/A
    & Always evict
    & N/A
    & Radix (LRU) \\
    & (2024) FastServe \cite{wu2024fast}
    & Paged (Tiered)
    & N/A
    & Always off.
    & Async (CPU to GPU)
    & N/A \\
Multi. mono.
    & (2024) Preble \cite{srivatsa2024preble}
    & Paged (GPU)
    & Hot entries
    & Always evict
    & N/A
    & Distr. radix (LRU) \\
Multi. disagg.
    & (2024) DistServe \cite{zhong2024distserve}
    & Paged (GPU)
    & N/A
    & N/A
    & Sync (P to D)
    & N/A \\
    & (2024) TetriInfer \cite{hu2024inference}
    & Paged (GPU)
    & N/A
    & N/A
    & Sync (P to D)
    & N/A \\
    & (2024) SplitWise \cite{patel2024splitwise}
    & Paged (GPU)
    & N/A
    & Always evict
    & Async (P to D)
    & N/A \\
    & (2024) Mooncake \cite{qin2024mooncake}
    & Paged (Tiered)
    & Hot entries
    & N/A
    & Async (P to D)
    & Hash table (LRU) \\
Multi. $\lambda$
    & (2025) DeepFlow \cite{hu2025deepflow}
    & Paged (Tiered)
    & N/A
    & Always evict
    & Async (CPU to GPU)
    & Radix (LRU) \\
\bottomrule
\end{tabular}
\end{table*}

An inference runtime can be targeted towards single-replica or multi-replica environments. For multi-replica runtimes specifically, since the workers housing each replica (distributed or not\footnote{Both single and multi-replica systems can support distributed execution, \textit{i.e.} model or pipeline parallelism, by storing model layers across multiple inference devices or machines. But in contrast to distributed database systems, there is no scope for optimizing the distributed execution plan, since layers cannot be executed out of sequence, nor any opportunity for optimizing the data partitioning, since no layer requires data from any other layer, apart from the intermediate results following a layer execution.}) may be under different loads, \textit{i.e.} processing requests towards various states of completion, with various memory burdens as a result of KV cache growth, along with shorter or longer request queues, multi-replica runtimes rely on load balancing techniques to selectively assign requests to replicas in order to maintain high performance. The ability to process multiple requests simultaneously, combined with blocked persistent caches, also leads to techniques for block replication and partitioning in order to take maximum advantage of cache reuse. Finally, the need to carefully manage the number of resource-intensive replicas leads to disaggregated and serverless architectures, in addition to monolithic architectures, that offer more granular scaling and control.

\subsubsection{Single-Replica}

Single-replica runtimes need to deal with fundamental system-level challenges for LLM execution, including batching, scheduling, and memory management. The techniques developed by these runtimes also serve as a foundation for multi-replica runtimes.

Orca \cite{yu2022orca} is one of the earliest inference runtimes specially designed to deal with the unique autoregressive characteristic of LLMs, pioneering the idea of round-based continuous batching that is now widespread. To manage cache memory, the runtime uses static memory allocation, preallocating a fixed amount of memory for every request based on maximum context length specified by the application. The scheduler does not allow for preemption, so there are no mechanisms for eviction or offloading preempted requests. As each memory block is private to each request, there is also no opportunity for cache sharing.

Other runtimes have been subsequently developed in order to address these various shortcomings. To address the limitations of static allocation, vLLM \cite{kwon2023efficient} introduces block-based cache memory. Memory blocks are handled by a memory manager, allowing for multiple active requests to share certain blocks. Even so, vLLM does not explicitly manage cache persistence across multiple request lifetimes. To respond to low-memory situations, vLLM adds support for request preemption based on request priority. Under low memory conditions, the scheduler evicts all cache blocks corresponding to the lowest priority request in order to free memory. Sarathi-Serve \cite{agrawal2024taming} extends the idea of continuous batching to the prefill stage, introducing chunked prefills on top of vLLM. Chunked prefills allows packing a batch with prefill tokens, allowing for dynamic batch sizing based on a token budget. SGLang \cite{zheng2024sglang} introduces managed cache persistence, developing a co-designed frontend and runtime to take advantage of this technique. The runtime persists all cache blocks in memory and uses a radix tree to speed up block retrieval during prefill. To control the size of the persisted cache, SGLang uses LRU eviction for persisted entries. To avoid cache thrashing, requests are prioritized based on length of shared prefix. For addressing the limitations of FCFS, FastServe \cite{wu2024fast} introduces MLQ job prioritization based on arrival time as a way of reducing average request latency.

\subsubsection{Multi-Replica}

\begin{figure*}[t]
\centering
\includegraphics[width=1.0\textwidth]{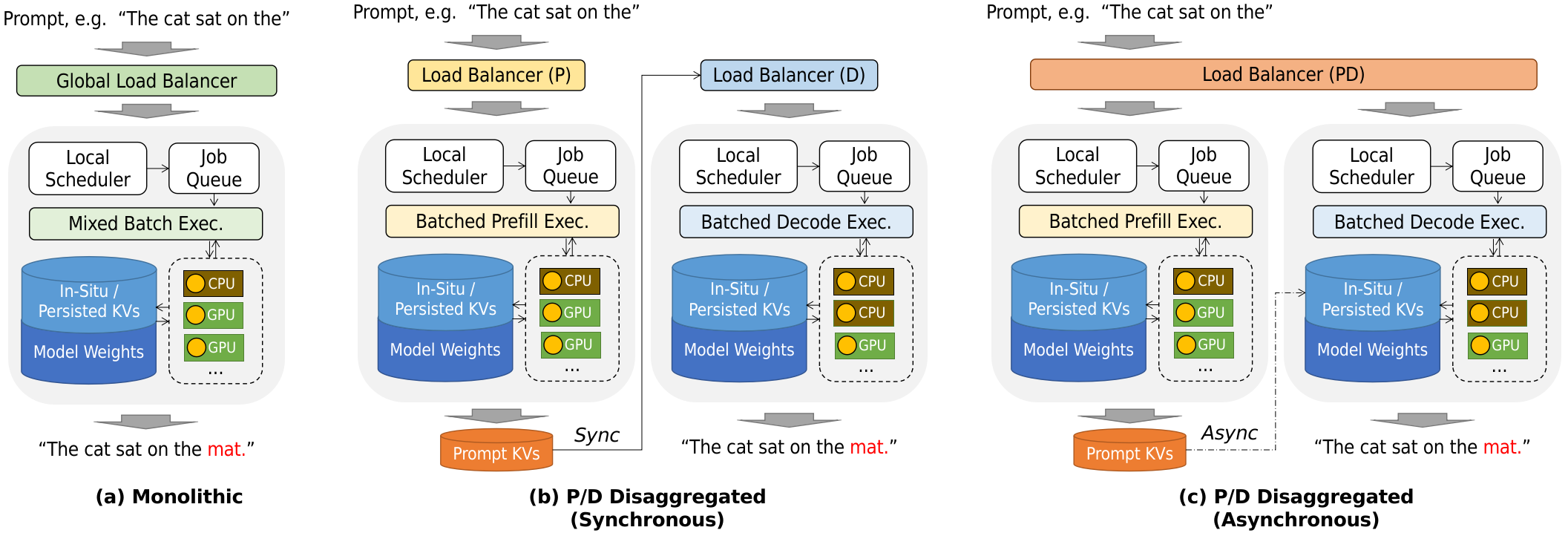}
\caption{Monolithic (non-disaggregated) (\textbf{a}) and P/D disaggregated (\textbf{b}, \textbf{c}) architectures.}
\label{fig:architectures}
\vspace{-1em}
\end{figure*}

Multi-replica runtimes use load balancers to evenly distribute the workload across multiple workers. For multi-replica runtimes that support cache persistent, the persisted entries can be distributed across different workers, leading to techniques for distributed cache management. Different application scaling needs have led to disaggregated and serverless runtimes, in addition to monolithic runtimes (Figure~\ref{fig:architectures}).

\hi{Monolithic.}
As an example of a monolithic runtime, \textit{Preble \cite{srivatsa2024preble}} demonstrates how cache persistence can be extended to a multi-replica scenario. A central scheduler maintains a global radix tree, and load balancing is performed based on cache hit ratio in addition to worker load. Since the persisted cache is distributed across the workers, some workers have higher hit rates. To avoid the load balancer overloading these workers, hot entries are replicated onto other workers.

\hi{Disaggregated.}
Compared to a monolithic architecture, using dedicated workers for the prefill and decode stages opens up opportunities for independent control and scaling of these two phases.

Before the decode worker can begin processing a request, the KV cache entries from the prefill must first be transferred onto the decode worker. For \textit{(1) synchronous transfer}, DistServe \cite{zhong2024distserve} reduces this transfer cost by carefully assigning inference devices (GPUs) in an inference cluster to either prefill tasks or decode tasks based on their physical locations in the cluster, in addition to needs of the workload. For example, by assigning GPUs in the same machine to independent prefill and decode tasks, the GPUs can take advantage of high-bandwidth NVLink to reduce this transfer cost. But since the decode worker is selected after the prefill stage (Figure~\ref{fig:architectures}(b)), if the workload suffers a burst, a greedy decode-side load balancer may inadvertantly overload a decode worker. TetriInfer \cite{hu2024inference} addresses this issue by adopting power-of-two load balancing \cite{mitzenmacher2001power}. For \textit{(2) asynchronous transfer}, SplitWise \cite{patel2024splitwise} and Mooncake \cite{qin2024mooncake} select the decode worker at the same time that the prefill worker is selected, allowing the cache to be streamed asynchronously onto the decode worker as it is being generated (Figure~\ref{fig:architectures}(c)). But since the load on the decode worker can change during prefill, the selected decode worker may no longer have enough available memory to process the request once prefill completes. To avoid preemptions, Mooncake checks the load on the decode worker before decode begins. If the memory is insufficient, the request is terminally aborted.

For hot entry replication, Mooncake uses a central hash table that catalogues the location and contents of each persisted cache block. During prefill, if a cache hit is detected, the reusable blocks are copied onto the prefill node, effectively replicating hot blocks across multiple workers.

\hi{Serverless.}
Serverless runtimes take advantage of stateless hardware resources to enable flexible resource scaling. But because the machines are stateless, these runtimes must overcome the problem of cold starts, where model weights must be loaded onto a machine at the time of procurement before it can be used.
For example, DeepFlow \cite{hu2025deepflow} is targeted at LLM inference over shared infrastructure. To avoid cold starts, it persists procured machines across multiple requests instead of returning them to the resource pool after each request while also keeping a prewarmed pool of machines for fast procurement. DeepFlow takes advantage of cache sharing opportunities by offloading cache entries onto host memory, managed by a central radix tree.

\subsection{Discussion}
\label{sec:systems-discussion}

While non-specialized deep learning inference systems like Clipper \cite{crankshaw2017clipper}, TensorFlow Serving \cite{olston2017tensorflowserving}, TensorRT, Clockwork \cite{gujarati2020serving}, and Hugging Face Accelerate\footnote{\url{http://huggingface.co/docs/accelerate}} can be applied to LLM inference, they are generally outperformed by dedicated LLM inference systems. This section presented a number of systems, each one introducing new innovations to increase throughput, reduce latency, and reduce memory usage, all while maintaining high quality inference. Many of these techniques can be used together without conflict, and indeed, recent system are already trending towards convergence. For example, techniques like paged attention and continuous batching with chunked prefills have already been widely adopted, and more recent techniques, such as MLQ scheduling and asynchronous cache recovery over tiered storage, may likewise become more widespread over time due to their clear advantages over prior approaches. In terms of system architecture, multi-replica systems include additional components such as load balancers and distributed cache management mechanisms, \textit{i.e.} hot cache replicas and cache transfer mechanisms. For single-replica environments, disabling these features in a multi-replica system can reduce the added overhead, otherwise a single-replica system may be more appropriate. For multi-replica environments, disaggregated systems allow for more flexible resource management without any significant downsides compared to monolithic systems, especially when using mixed worker pools. If deploying over shared infrastructure, serverless systems like DeepFlow \cite{hu2025deepflow} can be used to address issues including cold starts and stateless workers.

Quantization schemes may be one area where future systems can be differentiated. As mentioned in Section~\ref{sec:quantization}, there is an inherent tradeoff between accuracy and compression, and this tradeoff depends on the model architecture in addition to the task types that the requests entail. Hence, the optimal quantization scheme depends on the application and model that the inference system targets. On the other hand, there is also a clear need for inference systems that are more adaptive. Due to the unpredictable request lifecycles, workloads are inherently complex. While some of this unpredictability can be managed by constraining output generation, \textit{e.g.} permitting only constrained generation, or via workload disaggregation, where interactive requests that demand low latency are handled separately from non-interactive requests that benefit more from high throughput \cite{patke2025hierarchical,sun2025hygen}, LLM inference systems targeted at more general workloads will need more powerful techniques, including more accurate load prediction in addition to adaptive techniques such as dynamic batch sizing and elastic resource provisioning to handle workload shifts.

Efforts are already underway for developing more elastic inference systems. For example, for elastically scaling disaggregated systems, Splitwise \cite{patel2024splitwise} and TetriInfer \cite{hu2024inference} can dynamically adjust the ratio of prefill and decode workers depending on the workload. For static resource provisioning or for determining the initial resource configuration, iServe \cite{liakopoulos2025iserve} introduces the idea of lightweight fingerprint models that can be used to quickly assess different configurations.

\begin{figure}[t]
\centering
\includegraphics[width=0.48\textwidth]{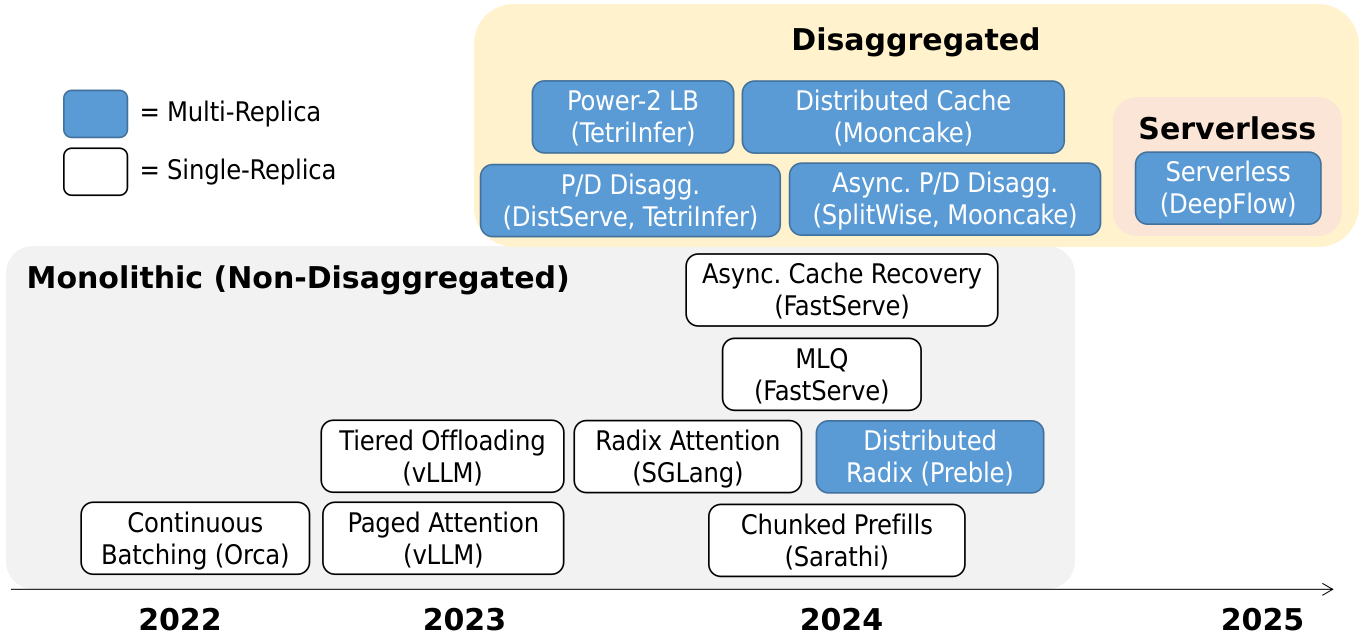}
\caption{Timeline of inference systems and key innovations.}
\label{fig:timeline}
\vspace{-1em}
\end{figure}

\section{Challenges and Open Problems}
\label{sec:open-problems}

High performance inference systems will become increasingly important as LLMs continue to be adopted for a variety of applications. For request processing operators and algorithms, the spread of LLMs beyond natural language \cite{luo2024projecting} could lead to new operators and sequence generation techniques due to the different attention patterns, opening up opportunities to develop more targeted techniques. For model execution, new developments in operators and algorithms naturally leads to new kernels, moreover load prediction remains an important problem for both job prioritization and load balancing. For memory management, the dizzying array of quantization schemes requires a more systematic understanding of when and what to quantize, in addition to more rigorous understanding of how to quantize. New logical memory hierarchies, such as pyramidal caches \cite{cai2024pyramidkv,yang2024pyramidinfer}, are also worth exploring further, in addition to novel techniques like entry merging \cite{liu2024minicache}. All these avenues motivate continued efforts for system design, especially towards developing elastic systems that can adaptively manage limited hardware resources in response to changes in the workload.

We mention some other opportunities for future work. Test-time scaling \cite{wu2024inference} focuses on delivering maximum inference quality within a fixed latency budget as opposed to directly minimizing latency, and consumer services like deep research\footnote{\url{http://gemini.google/overview/deep-research}} are already exploiting relaxed latency constraints for quality-sensitive applications like web analysis. Placing less emphasis on latency could lead to new inference system designs. For models that support profile switching via low-rank adapters (LoRA), adapter loading during inference also requires consideration \cite{gabrielsson2025compress,sheng2024slora}. Adding to this diversity is the rise of specialized LLMs, for example mobile LLMs targeted at resource-constrained devices \cite{zhao2024hetegen}, mutli-modal LLMs \cite{pan2024marconi}, and robotic LLMs for interacting with the physical world \cite{ling2024timelyllm}. Coupled with model design are advancements in hardware. The diverse characteristics of new devices adds additional complexity to job prioritization and load balancing, and some techniques have already emerged \cite{griggs2024melange,jiang2025demystifying}.

\section{Conclusion}

The rapid adoption of LLMs has led to the need for specialized high-performance inference systems to handle the high-velocity and high-volume workloads that LLM service providers now face,
moreover the autoregressive nature of LLM inference has led to the development of new techniques. In this survey, we showed that these techniques can be understood under a unified framework encompassing request processing, model execution, and memory management. To deal with the cost uncertainties posed by autoregressive generation, these techniques rely on load prediction, adaptive designs, and fundamental cost reduction approaches in order maintain high performance. These techniques culminate into single and multi-replica systems, including disaggregated systems which aim to offer even finer resource management via logical task disaggregation.

\bibliographystyle{abbrv}
\bibliography{main}

\begin{thebibliography}{100}

\bibitem{adnan2024keyformer}
M.~Adnan, A.~Arunkumar, G.~Jain, P.~Nair, I.~Soloveychik, and P.~Kamath.
\newblock Keyformer: {KV} cache reduction through key tokens selection for
  efficient generative inference.
\newblock In {\em MLSys'24}, volume~6, pages 114--127, 2024.

\bibitem{agrawal2024taming}
A.~Agrawal, N.~Kedia, A.~Panwar, J.~Mohan, N.~Kwatra, B.~Gulavani, A.~Tumanov,
  and R.~Ramjee.
\newblock Taming throughput-latency tradeoff in {LLM} inference with
  {Sarathi-Serve}.
\newblock In {\em OSDI'24}, pages 117--134, 2024.

\bibitem{ainslie2023gqa}
J.~Ainslie, J.~Lee-Thorp, M.~de~Jong, Y.~Zemlyanskiy, F.~Lebron, and
  S.~Sanghai.
\newblock {GQA}: {T}raining generalized multi-query transformer models from
  multi-head checkpoints.
\newblock In {\em EMNLP'23}, pages 4895--4901, 2023.

\bibitem{ainslie2020etc}
J.~Ainslie, S.~Ontanon, C.~Alberti, V.~Cvicek, Z.~Fisher, P.~Pham, A.~Ravula,
  S.~Sanghai, Q.~Wang, and L.~Yang.
\newblock {ETC}: Encoding long and structured inputs in transformers.
\newblock In {\em EMNLP'20}, pages 268--284, 2020.

\bibitem{aminabadi2022deepspeed}
R.~Y. Aminabadi, S.~Rajbhandari, A.~A. Awan, C.~Li, D.~Li, E.~Zheng, O.~Ruwase,
  S.~Smith, M.~Zhang, J.~Rasley, and Y.~He.
\newblock Deep{S}peed-{I}nference: {E}nabling efficient inference of
  transformer models at unprecedented scale.
\newblock In {\em SC'22}, 2022.

\bibitem{asai2023}
A.~Asai, S.~Min, Z.~Zhong, and D.~Chen.
\newblock Retrieval-based language models and applications.
\newblock In {\em ACL'23}, 2023.

\bibitem{beltagy2020longformer}
I.~Beltagy, M.~E. Peters, and A.~Cohan.
\newblock Longformer: The long-document transformer, 2020.

\bibitem{besta2024graph}
M.~Besta, N.~Blach, A.~Kubicek, R.~Gerstenberger, M.~Podstawski, L.~Gianinazzi,
  J.~Gajda, T.~Lehmann, H.~Niewiadomski, P.~Nyczyk, and T.~Hoefler.
\newblock Graph of thoughts: Solving elaborate problems with large language
  models.
\newblock {\em AAAI'24}, 38(16):17682--17690, 2024.

\bibitem{beurerkellner2023lmql}
L.~Beurer-Kellner, M.~Fischer, and M.~Vechev.
\newblock {LMQL Chat}: Scripted chatbot development, 2023.

\bibitem{beurerkellner2023prompting}
L.~Beurer-Kellner, M.~Fischer, and M.~Vechev.
\newblock Prompting is programming: A query language for large language models.
\newblock {\em Proc. ACM Program. Lang.}, 7, 2023.

\bibitem{boehm2023optimizing}
M.~Boehm, M.~Interlandi, and C.~Jermaine.
\newblock Optimizing tensor computations: {F}rom applications to compilation
  and runtime techniques.
\newblock In {\em SIGMOD'23}, pages 53--59, 2023.

\bibitem{bondarenko2021understanding}
Y.~Bondarenko, M.~Nagel, and T.~Blankevoort.
\newblock Understanding and overcoming the challenges of efficient transformer
  quantization.
\newblock In {\em EMNLP'21}, pages 7947--7969, 2021.

\bibitem{cai2024pyramidkv}
Z.~Cai, Y.~Zhang, B.~Gao, Y.~Liu, T.~Liu, K.~Lu, W.~Xiong, Y.~Dong, B.~Chang,
  J.~Hu, and W.~Xiao.
\newblock {PyramidKV}: Dynamic {KV} cache compression based on pyramidal
  information funneling, 2024.

\bibitem{cao2025locality}
S.~Cao, Y.~Wang, Z.~Mao, P.-L. Hsu, L.~Yin, T.~Xia, D.~Li, S.~Liu, Y.~Zhang,
  Y.~Zhou, Y.~Sheng, J.~Gonzalez, and I.~Stoica.
\newblock Locality-aware fair scheduling in {LLM} serving, 2025.

\bibitem{chavan2024faster}
A.~Chavan, R.~Magazine, S.~Kushwaha, M.~Debbah, and D.~Gupta.
\newblock Faster and lighter {LLMs}: {A} survey on current challenges and way
  forward.
\newblock In {\em IJCAI'24}, 2024.

\bibitem{chen2021evaluating}
M.~Chen, J.~Tworek, H.~Jun, Q.~Yuan, H.~P. de~Oliveira~Pinto, J.~Kaplan,
  H.~Edwards, Y.~Burda, N.~Joseph, G.~Brockman, A.~Ray, R.~Puri, G.~Krueger,
  M.~Petrov, H.~Khlaaf, G.~Sastry, P.~Mishkin, B.~Chan, S.~Gray, N.~Ryder,
  M.~Pavlov, A.~Power, L.~Kaiser, M.~Bavarian, C.~Winter, P.~Tillet, F.~P.
  Such, D.~Cummings, M.~Plappert, F.~Chantzis, E.~Barnes, A.~Herbert-Voss,
  W.~H. Guss, A.~Nichol, A.~Paino, N.~Tezak, J.~Tang, I.~Babuschkin, S.~Balaji,
  S.~Jain, W.~Saunders, C.~Hesse, A.~N. Carr, J.~Leike, J.~Achiam, V.~Misra,
  E.~Morikawa, A.~Radford, M.~Knight, M.~Brundage, M.~Murati, K.~Mayer,
  P.~Welinder, B.~McGrew, D.~Amodei, S.~McCandlish, I.~Sutskever, and
  W.~Zaremba.
\newblock Evaluating large language models trained on code, 2021.

\bibitem{chen1998evaluation}
S.~Chen, D.~Beeferman, and R.~Rosenfeld.
\newblock Evaluation metrics for language models.
\newblock Technical report, Carnegie Mellon University, 1998.

\bibitem{chen2024magic}
Z.~Chen, R.~Sadhukhan, Z.~Ye, Y.~Zhou, J.~Zhang, N.~Nolte, Y.~Tian, M.~Douze,
  L.~Bottou, Z.~Jia, and B.~Chen.
\newblock Magic{PIG}: {LSH} sampling for efficient {LLM} generation, 2024.

\bibitem{crankshaw2017clipper}
D.~Crankshaw, X.~Wang, G.~Zhou, M.~J. Franklin, J.~E. Gonzalez, and I.~Stoica.
\newblock Clipper: {A} low-latency online prediction serving system.
\newblock In {\em NSDI'17}, pages 613--627, 2017.

\bibitem{dao2022flashattention}
T.~Dao, D.~Y. Fu, S.~Ermon, A.~Rudra, and C.~R\'{e}.
\newblock Flash{A}ttention: {F}ast and memory-efficient exact attention with
  {IO}-awareness.
\newblock In {\em NeurIPS'22}, 2022.

\bibitem{deepseekai2025deepseek}
DeepSeek-AI.
\newblock {DeepSeek-R1}: Incentivizing reasoning capability in {LLM}s via
  reinforcement learning, 2025.

\bibitem{dettmers2022llm}
T.~Dettmers, M.~Lewis, Y.~Belkada, and L.~Zettlemoyer.
\newblock {LLM}.int8(): 8-bit matrix multiplication for transformers at scale.
\newblock In {\em NeurIPS'22}, 2022.

\bibitem{dettmers2023spqr}
T.~Dettmers, R.~Svirschevski, V.~Egiazarian, D.~Kuznedelev, E.~Frantar,
  S.~Ashkboos, A.~Borzunov, T.~Hoefler, and D.~Alistarh.
\newblock {SpQR}: A sparse-quantized representation for near-lossless {LLM}
  weight compression, 2023.

\bibitem{dong2024flex}
J.~Dong, B.~Feng, D.~Guessous, Y.~Liang, and H.~He.
\newblock Flex {A}ttention: A programming model for generating optimized
  attention kernels, 2024.

\bibitem{fang2021turbotransformers}
J.~Fang, Y.~Yu, C.~Zhao, and J.~Zhou.
\newblock Turbo{T}ransformers: {A}n efficient {GPU} serving system for
  transformer models.
\newblock In {\em PPoPP'21}, pages 389--402, 2021.

\bibitem{frantar2023optq}
E.~Frantar, S.~Ashkboos, T.~Hoefler, and D.~Alistarh.
\newblock {OPTQ}: {A}ccurate quantization for generative pre-trained
  transformers.
\newblock In {\em ICLR'23}, 2023.

\bibitem{freitag2017beam}
M.~Freitag and Y.~Al-Onaizan.
\newblock Beam search strategies for neural machine translation.
\newblock In {\em WNMT'17}, pages 56--60, 2017.

\bibitem{fu2024efficiently}
Y.~Fu, J.~Chen, S.~Zhu, Z.~Fu, Z.~Dai, A.~Qiao, and H.~Zhang.
\newblock Efficiently serving {LLM} reasoning programs with {C}ertaindex, 2024.

\bibitem{fu2024efficient}
Y.~Fu, S.~Zhu, R.~Su, A.~Qiao, I.~Stoica, and H.~Zhang.
\newblock Efficient {LLM} scheduling by learning to rank, 2024.

\bibitem{gabrielsson2025compress}
R.~B. Gabrielsson, J.~Zhu, O.~Bhardwaj, L.~Choshen, K.~Greenewald,
  M.~Yurochkin, and J.~Solomon.
\newblock Compress then serve: {S}erving thousands of {L}o{RA} adapters with
  little overhead, 2025.

\bibitem{gholami2021survey}
A.~Gholami, S.~Kim, Z.~Dong, Z.~Yao, M.~W. Mahoney, and K.~Keutzer.
\newblock A survey of quantization methods for efficient neural network
  inference, 2021.

\bibitem{gim2024prompt}
I.~Gim, G.~Chen, S.-s. Lee, N.~Sarda, A.~Khandelwal, and L.~Zhong.
\newblock {Prompt Cache}: Modular attention reuse for low-latency inference.
\newblock In {\em MLSys'24}, volume~6, pages 325--338, 2024.

\bibitem{graham1969bounds}
R.~L. Graham.
\newblock Bounds on multiprocessing timing anomalies.
\newblock {\em SIAM J. Appl. Math.}, 17(2):416--429, 1969.

\bibitem{graves2012sequence}
A.~Graves.
\newblock Sequence transduction with recurrent neural networks, 2012.

\bibitem{griggs2024melange}
T.~Griggs, X.~Liu, J.~Yu, D.~Kim, W.-L. Chiang, A.~Cheung, and I.~Stoica.
\newblock M\'elange: {C}ost efficient large language model serving by
  exploiting {GPU} heterogeneity, 2024.

\bibitem{gujarati2020serving}
A.~Gujarati, R.~Karimi, S.~Alzayat, W.~Hao, A.~Kaufmann, Y.~Vigfusson, and
  J.~Mace.
\newblock Serving {DNN}s like clockwork: {P}erformance predictability from the
  bottom up.
\newblock In {\em OSDI'20}, pages 443--462, 2020.

\bibitem{holtzman2020curious}
A.~Holtzman, J.~Buys, L.~Du, M.~Forbes, and Y.~Choi.
\newblock The curious case of neural text degeneration, 2020.

\bibitem{hong2024flashdecoding}
K.~Hong, G.~Dai, J.~Xu, Q.~Mao, X.~Li, J.~Liu, k.~chen, Y.~Dong, and Y.~Wang.
\newblock {FlashDecoding++}: Faster large language model inference with
  asynchronization, flat {GeMM} optimization, and heuristics.
\newblock In {\em MLSys'24}, volume~6, pages 148--161, 2024.

\bibitem{hu2024inference}
C.~Hu, H.~Huang, L.~Xu, X.~Chen, J.~Xu, S.~Chen, H.~Feng, C.~Wang, S.~Wang,
  Y.~Bao, N.~Sun, and Y.~Shan.
\newblock Inference without interference: Disaggregate {LLM} inference for
  mixed downstream workloads, 2024.

\bibitem{hu2025epic}
J.~Hu, W.~Huang, H.~Wang, W.~Wang, T.~Hu, Q.~Zhang, H.~Feng, X.~Chen, Y.~Shan,
  and T.~Xie.
\newblock E{PIC}: {E}fficient position-independent context caching for serving
  large language models, 2025.

\bibitem{hu2025deepflow}
J.~Hu, J.~Xu, Z.~Liu, Y.~He, Y.~Chen, H.~Xu, J.~Liu, B.~Zhang, S.~Wan, G.~Dan,
  Z.~Dong, Z.~Ren, J.~Meng, C.~He, C.~Liu, T.~Xie, D.~Lin, Q.~Zhang, Y.~Yu,
  H.~Feng, X.~Chen, and Y.~Shan.
\newblock {DeepFlow}: Serverless large language model serving at scale, 2025.

\bibitem{jiang2024minference}
H.~Jiang, Y.~Li, C.~Zhang, Q.~Wu, X.~Luo, S.~Ahn, Z.~Han, A.~H. Abdi, D.~Li,
  C.-Y. Lin, Y.~Yang, and L.~Qiu.
\newblock {MInference} 1.0: Accelerating pre-filling for long-context {LLMS}
  via dynamic sparse attention, 2024.

\bibitem{jiang2025demystifying}
Y.~Jiang, F.~Fu, X.~Yao, G.~He, X.~Miao, A.~Klimovic, B.~Cui, B.~Yuan, and
  E.~Yoneki.
\newblock Demystifying cost-efficiency in {LLM} serving over heterogeneous
  {GPU}s, 2025.

\bibitem{jin2023s3}
Y.~Jin, C.-F. Wu, D.~Brooks, and G.-Y. Wei.
\newblock S$^3$: Increasing {GPU} utilization during generative inference for
  higher throughput.
\newblock In {\em NeurIPS'23}, 2023.

\bibitem{juravsky2024hydragen}
J.~Juravsky, B.~Brown, R.~Ehrlich, D.~Y. Fu, C.~Ré, and A.~Mirhoseini.
\newblock Hydragen: High-throughput {LLM} inference with shared prefixes, 2024.

\bibitem{khattab2024dspy}
O.~Khattab, A.~Singhvi, P.~Maheshwari, Z.~Zhang, K.~Santhanam, S.~Vardhamanan,
  S.~Haq, A.~Sharma, T.~T. Joshi, H.~Moazam, H.~Miller, M.~Zaharia, and
  C.~Potts.
\newblock {DSPy}: Compiling declarative language model calls into
  self-improving pipelines.
\newblock In {\em ICLR'24}, 2024.

\bibitem{khoshnoodi2024comprehensive}
M.~Khoshnoodi, V.~Jain, M.~Gao, M.~Srikanth, and A.~Chadha.
\newblock A comprehensive survey of accelerated generation techniques in large
  language models, 2024.

\bibitem{kim2024trustworthy}
K.~Kim and A.~Ailamaki.
\newblock Trustworthy and efficient {LLM}s meet databases, 2024.

\bibitem{kossmann2025gpu}
F.~Kossmann, B.~Fontaine, D.~Khudia, M.~Cafarella, and S.~Madden.
\newblock Is the {GPU} half-empty or half-full? {P}ractical scheduling
  techniques for {LLM}s, 2025.

\bibitem{kwon2025lol}
H.~Kwon, K.~Koo, J.~Kim, W.~Lee, M.~Lee, H.~Lee, Y.~Jung, J.~Park, Y.~Song,
  B.~Yang, H.~Choi, G.~Kim, J.~Won, W.~Shin, C.~Kim, G.~Shin, Y.~Kwon, I.~Kim,
  E.~Lim, J.~Kim, and J.~Choi.
\newblock Lo{L}-{PIM}: {L}ong-context {LLM} decoding with scalable {DRAM-PIM}
  system, 2025.

\bibitem{kwon2023efficient}
W.~Kwon, Z.~Li, S.~Zhuang, Y.~Sheng, L.~Zheng, C.~H. Yu, J.~Gonzalez, H.~Zhang,
  and I.~Stoica.
\newblock Efficient memory management for large language model serving with
  {PagedAttention}.
\newblock In {\em SOSP'23}, pages 611--626, 2023.

\bibitem{lee2024infinigen}
W.~Lee, J.~Lee, J.~Seo, and J.~Sim.
\newblock {InfiniGen}: {E}fficient generative inference of large language
  models with dynamic {KV} cache management.
\newblock In {\em OSDI'24}, pages 155--172, 2024.

\bibitem{leviathan2023fast}
Y.~Leviathan, M.~Kalman, and Y.~Matias.
\newblock Fast inference from transformers via speculative decoding.
\newblock In {\em ICML'23}, 2023.

\bibitem{li2024llm}
B.~Li, Y.~Jiang, V.~Gadepally, and D.~Tiwari.
\newblock {LLM} inference serving: {S}urvey of recent advances and
  opportunities, 2024.

\bibitem{li2025survey}
H.~Li, Y.~Li, A.~Tian, T.~Tang, Z.~Xu, X.~Chen, N.~Hu, W.~Dong, Q.~Li, and
  L.~Chen.
\newblock A survey on large language model acceleration based on {KV} cache
  management, 2025.

\bibitem{liakopoulos2025iserve}
D.~Liakopoulos, T.~Hu, P.~Sinha, and N.~J. Yadwadkar.
\newblock i{S}erve: An intent-based serving system for {LLM}s, 2025.

\bibitem{lin2004rouge}
C.-Y. Lin.
\newblock {ROUGE}: A package for automatic evaluation of summaries.
\newblock In {\em ACL'04}, pages 74--81, 2004.

\bibitem{lin2025qserve}
Y.~Lin, H.~Tang, S.~Yang, Z.~Zhang, G.~Xiao, C.~Gan, and S.~Han.
\newblock {QServe}: {W4A8KV4} quantization and system co-design for efficient
  {LLM} serving.
\newblock In {\em MLSys'25}, 2025.

\bibitem{ling2024timelyllm}
N.~Ling, G.~Chen, and L.~Zhong.
\newblock Timely{LLM}: {S}egmented {LLM} serving system for time-sensitive
  robotic applications, 2024.

\bibitem{liu2024minicache}
A.~Liu, J.~Liu, Z.~Pan, Y.~He, G.~Haffari, and B.~Zhuang.
\newblock {MiniCache}: {KV} cache compression in depth dimension for large
  language models, 2024.

\bibitem{liu2024retrieval}
D.~Liu, M.~Chen, B.~Lu, H.~Jiang, Z.~Han, Q.~Zhang, Q.~Chen, C.~Zhang, B.~Ding,
  K.~Zhang, C.~Chen, F.~Yang, Y.~Yang, and L.~Qiu.
\newblock Retrieval{A}ttention: Accelerating long-context {LLM} inference via
  vector retrieval.
\newblock In {\em NeurIPS'24}, 2024.

\bibitem{liu2023blockwise}
H.~Liu and P.~Abbeel.
\newblock Blockwise parallel transformers for large context models.
\newblock In {\em NeurIPS'23}, volume~36, pages 8828--8844, 2023.

\bibitem{liu2023ring}
H.~Liu, M.~Zaharia, and P.~Abbeel.
\newblock Ring attention with blockwise transformers for near-infinite context,
  2023.

\bibitem{liu2025survey}
J.~Liu, P.~Tang, W.~Wang, Y.~Ren, X.~Hou, P.-A. Heng, M.~Guo, and C.~Li.
\newblock A survey on inference optimization techniques for mixture of experts
  models, 2025.

\bibitem{liu2025mell}
Q.~Liu, Z.~Hong, F.~Chen, P.~Li, and S.~Guo.
\newblock Mell: Memory-efficient large language model serving via multi-{GPU}
  {KV} cache management, 2025.

\bibitem{liu2024optimizing}
S.~Liu, A.~Biswal, A.~Cheng, X.~Mo, S.~Cao, J.~E. Gonzalez, I.~Stoica, and
  M.~Zaharia.
\newblock Optimizing {LLM} queries in relational workloads, 2024.

\bibitem{liu2023scissorhands}
Z.~Liu, A.~Desai, F.~Liao, W.~Wang, V.~Xie, Z.~Xu, A.~Kyrillidis, and
  A.~Shrivastava.
\newblock Scissorhands: Exploiting the persistence of importance hypothesis for
  {LLM} {KV} cache compression at test time.
\newblock In {\em NeurIPS'23}, 2023.

\bibitem{luo2024projecting}
S.~Luo, W.~Gao, Z.~Wu, J.~Peng, C.~W. Coley, and J.~Ma.
\newblock Projecting molecules into synthesizable chemical spaces, 2024.

\bibitem{milakov2018online}
M.~Milakov and N.~Gimelshein.
\newblock Online normalizer calculation for softmax, 2018.

\bibitem{mitzenmacher2001power}
M.~Mitzenmacher.
\newblock The power of two choices in randomized load balancing.
\newblock {\em IEEE Trans. Parallel and Distrib. Syst.}, 12(10):1094--1104,
  2001.

\bibitem{nagel2021white}
M.~Nagel, M.~Fournarakis, R.~A. Amjad, Y.~Bondarenko, M.~van Baalen, and
  T.~Blankevoort.
\newblock A white paper on neural network quantization, 2021.

\bibitem{narayan2022can}
A.~Narayan, I.~Chami, L.~Orr, and C.~R\'{e}.
\newblock Can foundation models wrangle your data?
\newblock {\em Proc. VLDB Endow.}, 16(4):738--746, 2022.

\bibitem{olston2017tensorflowserving}
C.~Olston, N.~Fiedel, K.~Gorovoy, J.~Harmsen, L.~Lao, F.~Li, V.~Rajashekhar,
  S.~Ramesh, and J.~Soyke.
\newblock {T}ensor{F}low-{S}erving: {F}lexible, high-performance {ML} serving.
\newblock In {\em NeurIPS'17 Workshop on ML Systems}, 2017.

\bibitem{oren2024transformers}
M.~Oren, M.~Hassid, N.~Yarden, Y.~Adi, and R.~Schwartz.
\newblock Transformers are multi-state {RNN}s, 2024.

\bibitem{osama2023streamk}
M.~Osama, D.~Merrill, C.~Cecka, M.~Garland, and J.~D. Owens.
\newblock {Stream-K}: Work-centric parallel decomposition for dense
  matrix-matrix multiplication on the {GPU}, 2023.

\bibitem{pan2024marconi}
R.~Pan, Z.~Wang, Z.~Jia, C.~Karakus, L.~Zancato, T.~Dao, Y.~Wang, and
  R.~Netravali.
\newblock Marconi: {P}refix caching for the era of hybrid {LLM}s, 2024.

\bibitem{papineni2002bleu}
K.~Papineni, S.~Roukos, T.~Ward, and W.-J. Zhu.
\newblock {BLEU}: A method for automatic evaluation of machine translation.
\newblock In {\em ACL'02}, pages 311--318, 2002.

\bibitem{patel2024splitwise}
P.~Patel, E.~Choukse, C.~Zhang, A.~Shah, I.~Goiri, S.~Maleki, and R.~Bianchini.
\newblock Splitwise: {E}fficient generative {LLM} inference using phase
  splitting.
\newblock In {\em ISCA'24}, pages 118--132, 2024.

\bibitem{patke2025hierarchical}
A.~Patke, D.~Reddy, S.~Jha, C.~Narayanaswami, Z.~Kalbarczyk, and R.~Iyer.
\newblock Hierarchical autoscaling for large language model serving with
  {Chiron}, 2025.

\bibitem{peeperkorn2024temperature}
M.~Peeperkorn, T.~Kouwenhoven, D.~Brown, and A.~Jordanous.
\newblock Is temperature the creativity parameter of large language models?,
  2024.

\bibitem{prabhu2025vattention}
R.~Prabhu, A.~Nayak, J.~Mohan, R.~Ramjee, and A.~Panwar.
\newblock v{A}ttention: {D}ynamic memory management for serving {LLM}s without
  {P}aged{A}ttention, 2025.

\bibitem{qin2024mooncake}
R.~Qin, Z.~Li, W.~He, M.~Zhang, Y.~Wu, W.~Zheng, and X.~Xu.
\newblock Mooncake: A {KVC}ache-centric disaggregated architecture for {LLM}
  serving, 2024.

\bibitem{ramapuram2025}
J.~Ramapuram, F.~Danieli, E.~Dhekane, F.~Weers, D.~Busbridge, P.~Ablin,
  T.~Likhomanenko, J.~Digani, Z.~Gu, A.~Shidani, and R.~Webb.
\newblock Theory, analysis, and best practices for sigmoid self-attention,
  2025.

\bibitem{ren2024efficacy}
S.~Ren and K.~Q. Zhu.
\newblock On the efficacy of eviction policy for key-value constrained
  generative language model inference, 2024.

\bibitem{ribar2024sparq}
L.~Ribar, I.~Chelombiev, L.~Hudlass-Galley, C.~Blake, C.~Luschi, and D.~Orr.
\newblock {SparQ} attention: Bandwidth-efficient {LLM} inference.
\newblock In {\em Forty-first International Conference on Machine Learning},
  2024.

\bibitem{sahoo2025systematic}
P.~Sahoo, A.~K. Singh, S.~Saha, V.~Jain, S.~Mondal, and A.~Chadha.
\newblock A systematic survey of prompt engineering in large language models:
  {T}echniques and applications, 2025.

\bibitem{sanh2020distilbert}
V.~Sanh, L.~Debut, J.~Chaumond, and T.~Wolf.
\newblock {DistilBERT}, a distilled version of {BERT}: Smaller, faster, cheaper
  and lighter, 2020.

\bibitem{sanovar2025lean}
R.~Sanovar, S.~Bharadwaj, R.~S. Amant, V.~Rühle, and S.~Rajmohan.
\newblock Lean attention: Hardware-aware scalable attention mechanism for the
  decode-phase of transformers, 2025.

\bibitem{shahout2024dont}
R.~Shahout, E.~Malach, C.~Liu, W.~Jiang, M.~Yu, and M.~Mitzenmacher.
\newblock Don't stop me now: Embedding based scheduling for {LLM}s, 2024.

\bibitem{shazeer2019fast}
N.~Shazeer.
\newblock Fast transformer decoding: One write-head is all you need, 2019.

\bibitem{shazeer2017}
N.~Shazeer, A.~Mirhoseini, K.~Maziarz, A.~Davis, Q.~Le, G.~Hinton, and J.~Dean.
\newblock Outrageously large neural networks: The sparsely-gated
  mixture-of-experts layer.
\newblock In {\em ICLR'17}, 2017.

\bibitem{sheng2024slora}
Y.~Sheng, S.~Cao, D.~Li, C.~Hooper, N.~Lee, S.~Yang, C.~Chou, B.~Zhu, L.~Zheng,
  K.~Keutzer, J.~E. Gonzalez, and I.~Stoica.
\newblock {S-LoRA}: {S}erving thousands of concurrent {LoRA} adapters.
\newblock In {\em MLSys'24}, 2023.

\bibitem{sheng2023flexgen}
Y.~Sheng, L.~Zheng, B.~Yuan, Z.~Li, M.~Ryabinin, B.~Chen, P.~Liang, C.~R\'{e},
  I.~Stoica, and C.~Zhang.
\newblock Flex{G}en: {H}igh-throughput generative inference of large language
  models with a single {GPU}.
\newblock In {\em ICML'23}, 2023.

\bibitem{srivatsa2024preble}
V.~Srivatsa, Z.~He, R.~Abhyankar, D.~Li, and Y.~Zhang.
\newblock Preble: Efficient distributed prompt scheduling for {LLM} serving,
  2024.

\bibitem{sun2024llumnix}
B.~Sun, Z.~Huang, H.~Zhao, W.~Xiao, X.~Zhang, Y.~Li, and W.~Lin.
\newblock Llumnix: {D}ynamic scheduling for large language model serving.
\newblock In {\em OSDI'24}, pages 173--191, 2024.

\bibitem{sun2025hygen}
T.~Sun, P.~Wang, and F.~Lai.
\newblock Hy{G}en: {E}fficient {LLM} serving via elastic online-offline request
  co-location, 2025.

\bibitem{vaswani2017attention}
A.~Vaswani, N.~Shazeer, N.~Parmar, J.~Uszkoreit, L.~Jones, A.~N. Gomez,
  L.~Kaiser, and I.~Polosukhin.
\newblock Attention is all you need.
\newblock In {\em NeurIPS'17}, pages 6000--6010, 2017.

\bibitem{wan2024d2o}
Z.~Wan, X.~Wu, Y.~Zhang, Y.~Xin, C.~Tao, Z.~Zhu, X.~Wang, S.~Luo, J.~Xiong, and
  M.~Zhang.
\newblock {D2O}: Dynamic discriminative operations for efficient generative
  inference of large language models, 2024.

\bibitem{wang2024flash}
G.~Wang, J.~Zeng, X.~Xiao, S.~Wu, J.~Yang, L.~Zheng, Z.~Chen, J.~Bian, D.~Yu,
  and H.~Wang.
\newblock {FlashMask}: Efficient and rich mask extension of {FlashAttention},
  2024.

\bibitem{wang2025aop}
J.~Wang and G.~Li.
\newblock {AOP}: Automated and interactive {LLM} pipeline orchestration for
  answering complex queries.
\newblock In {\em CIDR'25}, 2025.

\bibitem{wang2023self}
X.~Wang, J.~Wei, D.~Schuurmans, Q.~Le, E.~Chi, S.~Narang, A.~Chowdhery, and
  D.~Zhou.
\newblock Self-consistency improves chain of thought reasoning in language
  models.
\newblock In {\em ICLR'23}, 2023.

\bibitem{wang2021lightseq}
X.~Wang, Y.~Xiong, Y.~Wei, M.~Wang, and L.~Li.
\newblock {L}ight{S}eq: {A} high performance inference library for
  transformers.
\newblock In {\em NAACL'21}, pages 113--120, 2021.

\bibitem{wang2024model}
Z.~Wang, B.~Jin, Z.~Yu, and M.~Zhang.
\newblock Model tells you where to merge: Adaptive {KV} cache merging for
  {LLM}s on long-context tasks, 2024.

\bibitem{wang2024revisiting}
Z.~Wang, S.~Li, Y.~Zhou, X.~Li, R.~Gu, N.~Cam-Tu, C.~Tian, and S.~Zhong.
\newblock Revisiting {SLO} and goodput metrics in {LLM} serving, 2024.

\bibitem{wei2022chain}
J.~Wei, X.~Wang, D.~Schuurmans, M.~Bosma, b.~ichter, F.~Xia, E.~Chi, Q.~V. Le,
  and D.~Zhou.
\newblock Chain-of-thought prompting elicits reasoning in large language
  models.
\newblock In {\em NeurIPS'22}, volume~35, pages 24824--24837, 2022.

\bibitem{welleck2019neural}
S.~Welleck, I.~Kulikov, S.~Roller, E.~Dinan, K.~Cho, and J.~Weston.
\newblock Neural text generation with unlikelihood training, 2019.

\bibitem{willard2023efficient}
B.~T. Willard and R.~Louf.
\newblock Efficient guided generation for large language models, 2023.

\bibitem{wu2024fast}
B.~Wu, Y.~Zhong, Z.~Zhang, S.~Liu, F.~Liu, Y.~Sun, G.~Huang, X.~Liu, and
  X.~Jin.
\newblock Fast distributed inference serving for large language models, 2024.

\bibitem{wu2024inference}
Y.~Wu, Z.~Sun, S.~Li, S.~Welleck, and Y.~Yang.
\newblock Inference scaling laws: {A}n empirical analysis of compute-optimal
  inference for problem-solving with language models, 2024.

\bibitem{xia2024unlocking}
H.~Xia, Z.~Yang, Q.~Dong, P.~Wang, Y.~Li, T.~Ge, T.~Liu, W.~Li, and Z.~Sui.
\newblock Unlocking efficiency in large language model inference: {A}
  comprehensive survey of speculative decoding, 2024.

\bibitem{xiao2024infllm}
C.~Xiao, P.~Zhang, X.~Han, G.~Xiao, Y.~Lin, Z.~Zhang, Z.~Liu, and M.~Sun.
\newblock {InfLLM}: Training-free long-context extrapolation for {LLM}s with an
  efficient context memory, 2024.

\bibitem{xiao2023smoothquant}
G.~Xiao, J.~Lin, M.~Seznec, H.~Wu, J.~Demouth, and S.~Han.
\newblock {SmoothQuant}: Accurate and efficient post-training quantization for
  large language models.
\newblock In {\em ICML'23}, 2023.

\bibitem{xiao2024duoattention}
G.~Xiao, J.~Tang, J.~Zuo, J.~Guo, S.~Yang, H.~Tang, Y.~Fu, and S.~Han.
\newblock {DuoAttention}: Efficient long-context {LLM} inference with retrieval
  and streaming heads, 2024.

\bibitem{xiao2024efficient}
G.~Xiao, Y.~Tian, B.~Chen, S.~Han, and M.~Lewis.
\newblock Efficient streaming language models with attention sinks.
\newblock In {\em ICLR'24}, 2024.

\bibitem{xu2024does}
D.~Xu, T.~Xie, B.~Xia, H.~Li, Y.~Bai, Y.~Sun, and W.~Wang.
\newblock Does few-shot learning help {LLM} performance in code synthesis?,
  2024.

\bibitem{xu2025resource}
M.~Xu, D.~Cai, W.~Yin, S.~Wang, X.~Jin, and X.~Liu.
\newblock Resource-efficient algorithms and systems of foundation models: {A}
  survey.
\newblock {\em ACM Comput. Surv.}, 57(5), 2025.

\bibitem{zhao2024hetegen}
Z.~Xuanlei, B.~Jia, H.~Zhou, Z.~Liu, S.~Cheng, and Y.~You.
\newblock Hete{G}en: {E}fficient heterogeneous parallel inference for large
  language models on resource-constrained devices.
\newblock In {\em MLSys'24}, volume~6, pages 162--172, 2024.

\bibitem{yan2025decoding}
M.~Yan, S.~Agarwal, and S.~Venkataraman.
\newblock Decoding speculative decoding, 2025.

\bibitem{yan2021fastseq}
Y.~Yan, F.~Hu, J.~Chen, N.~Bhendawade, T.~Ye, Y.~Gong, N.~Duan, D.~Cui, B.~Chi,
  and R.~Zhang.
\newblock {F}ast{S}eq: {M}ake sequence generation faster.
\newblock In H.~Ji, J.~C. Park, and R.~Xia, editors, {\em ACL-IJCNLP'21}, pages
  218--226, 2021.

\bibitem{yang2024pyramidinfer}
D.~Yang, X.~Han, Y.~Gao, Y.~Hu, S.~Zhang, and H.~Zhao.
\newblock {PyramidInfer}: Pyramid {KV} cache compression for high-throughput
  {LLM} inference.
\newblock In {\em ACL'24}, pages 3258--3270, 2024.

\bibitem{yao2024cacheblend}
J.~Yao, H.~Li, Y.~Liu, S.~Ray, Y.~Cheng, Q.~Zhang, K.~Du, S.~Lu, and J.~Jiang.
\newblock {CacheBlend}: Fast large language model serving for {RAG} with cached
  knowledge fusion, 2024.

\bibitem{yao2023tree}
S.~Yao, D.~Yu, J.~Zhao, I.~Shafran, T.~L. Griffiths, Y.~Cao, and K.~Narasimhan.
\newblock Tree of thoughts: Deliberate problem solving with large language
  models.
\newblock In {\em NeurIPS'23}, 2023.

\bibitem{yao2022zeroquant}
Z.~Yao, R.~Y. Aminabadi, M.~Zhang, X.~Wu, C.~Li, and Y.~He.
\newblock {ZeroQuant}: Efficient and affordable post-training quantization for
  large-scale transformers.
\newblock In {\em NeurIPS'22}, 2022.

\bibitem{ye2024chunkattention}
L.~Ye, Z.~Tao, Y.~Huang, and Y.~Li.
\newblock {ChunkAttention}: Efficient self-attention with prefix-aware {KV}
  cache and two-phase partition.
\newblock In {\em ACL'24}, pages 11608--11620, 2024.

\bibitem{ye2025flashinfer}
Z.~Ye, L.~Chen, R.~Lai, W.~Lin, Y.~Zhang, S.~Wang, T.~Chen, B.~Kasikci,
  V.~Grover, A.~Krishnamurthy, and L.~Ceze.
\newblock Flash{I}nfer: {E}fficient and customizable attention engine for {LLM}
  inference serving, 2025.

\bibitem{yu2022orca}
G.-I. Yu, J.~S. Jeong, G.-W. Kim, S.~Kim, and B.-G. Chun.
\newblock Orca: {A} distributed serving system for transformer-based generative
  models.
\newblock In {\em OSDI'22}, pages 521--538, 2022.

\bibitem{zadeh2020gobo}
A.~H. Zadeh, I.~Edo, O.~M. Awad, and A.~Moshovos.
\newblock {GOBO}: Quantizing attention-based {NLP} models for low latency and
  energy efficient inference.
\newblock In {\em MICRO'20}, pages 811--824, 2020.

\bibitem{zeng2022boosting}
J.~Zeng, M.~Li, Z.~Wu, J.~Liu, Y.~Liu, D.~Yu, and Y.~Ma.
\newblock Boosting distributed training performance of the unpadded {BERT}
  model, 2022.

\bibitem{zhai2023bytetransformer}
Y.~Zhai, C.~Jiang, L.~Wang, X.~Jia, S.~Zhang, Z.~Chen, X.~Liu, and Y.~Zhu.
\newblock Byte{T}ransformer: {A} high-performance transformer boosted for
  variable-length inputs.
\newblock In {\em IPDPS'23}, pages 344--355, 2023.

\bibitem{zhang2024pqcache}
H.~Zhang, X.~Ji, Y.~Chen, F.~Fu, X.~Miao, X.~Nie, W.~Chen, and B.~Cui.
\newblock {PQCache}: Product quantization-based {KVC}ache for long context
  {LLM} inference, 2024.

\bibitem{zhang2025sageattention}
J.~Zhang, J.~wei, P.~Zhang, J.~Zhu, and J.~Chen.
\newblock {SageAttention}: Accurate 8-bit attention for plug-and-play inference
  acceleration.
\newblock In {\em ICML'25}, 2025.

\bibitem{zhang2022opt}
S.~Zhang, S.~Roller, N.~Goyal, M.~Artetxe, M.~Chen, S.~Chen, C.~Dewan, M.~Diab,
  X.~Li, X.~V. Lin, T.~Mihaylov, M.~Ott, S.~Shleifer, K.~Shuster, D.~Simig,
  P.~S. Koura, A.~Sridhar, T.~Wang, and L.~Zettlemoyer.
\newblock {OPT}: Open pre-trained transformer language models, 2022.

\bibitem{zhang2024cam}
Y.~Zhang, Y.~Du, G.~Luo, Y.~Zhong, Z.~Zhang, S.~Liu, and R.~Ji.
\newblock {CaM}: Cache merging for memory-efficient {LLM}s inference.
\newblock In {\em PMLR'24}, volume 235, pages 58840--58850, 2024.

\bibitem{zhang2024reactable}
Y.~Zhang, J.~Henkel, A.~Floratou, J.~Cahoon, S.~Deep, and J.~M. Patel.
\newblock {ReAcTable}: Enhancing {ReAct} for table question answering.
\newblock {\em Proc. VLDB Endow.}, 17(8):1981--4, 2024.

\bibitem{zhang2023h2o}
Z.~Zhang, Y.~Sheng, T.~Zhou, T.~Chen, L.~Zheng, R.~Cai, Z.~Song, Y.~Tian,
  C.~R\'{e}, C.~Barrett, Z.~Wang, and B.~Chen.
\newblock H2{O}: {H}eavy-hitter oracle for efficient generative inference of
  large language models.
\newblock In {\em NeurIPS'23}, 2023.

\bibitem{zheng2024sglang}
L.~Zheng, L.~Yin, Z.~Xie, C.~Sun, J.~Huang, C.~H. Yu, S.~Cao, C.~Kozyrakis,
  I.~Stoica, J.~E. Gonzalez, C.~Barrett, and Y.~Sheng.
\newblock {SGLang}: Efficient execution of structured language model programs,
  2024.

\bibitem{zheng2023response}
Z.~Zheng, X.~Ren, F.~Xue, Y.~Luo, X.~Jiang, and Y.~You.
\newblock Response length perception and sequence scheduling: An
  {LLM}-empowered {LLM} inference pipeline.
\newblock In {\em NeurIPS'23}, 2023.

\bibitem{zhong2024distserve}
Y.~Zhong, S.~Liu, J.~Chen, J.~Hu, Y.~Zhu, X.~Liu, X.~Jin, and H.~Zhang.
\newblock {DistServe}: Disaggregating prefill and decoding for
  goodput-optimized large language model serving.
\newblock In {\em OSDI'24}, pages 193--210, 2024.

\bibitem{zhou2024survey}
Z.~Zhou, X.~Ning, K.~Hong, T.~Fu, J.~Xu, S.~Li, Y.~Lou, L.~Wang, Z.~Yuan,
  X.~Li, S.~Yan, G.~Dai, X.-P. Zhang, Y.~Dong, and Y.~Wang.
\newblock A survey on efficient inference for large language models, 2024.

\bibitem{zhu2024relay}
L.~Zhu, X.~Wang, W.~Zhang, and R.~W.~H. Lau.
\newblock Relay{A}ttention for efficient large language model serving with long
  system prompts, 2024.

\end{thebibliography}

\end{document}